\documentclass[aip,pof,amsmath,amssymb,preprint,linenumbers,
superscriptaddress,
]{revtex4-1}
\ifx\pdfoutput\undefined
\usepackage{graphicx}
\else
\usepackage{graphicx}
\usepackage{epstopdf}
\fi
\usepackage[pdftitle={PoF},bookmarks=true,citecolor=red,colorlinks=false]{hyperref}
\usepackage{wasysym}
\usepackage{xcolor}

\usepackage{lineno}

\usepackage{CJK}
\newcommand{\upd}{\mathrm{\,d}}

\newcommand{\red}[1]{\textcolor{black}{#1}}

\begin{document}

  \linenumbers
\preprint{AIP/PoF}
\begin{CJK*}{GB}{gbsn} 
\title{Hilbert  Statistics of Vorticity Scaling  in Two-Dimensional Turbulence}

\author{H.S. Tan (Ì·»½Êé)}
\affiliation{ Shanghai Institute of Applied Mathematics and Mechanics, Shanghai Key Laboratory of Mechanics in Energy Engineering, Shanghai University,
Shanghai 200072, People\rq{}s Republic of China
}

\author{Y.X. Huang (»ÆÓÀÏé)}
\email{yongxianghuang@gmail.com}
\affiliation{ Shanghai Institute of Applied Mathematics and Mechanics, Shanghai Key Laboratory of Mechanics in Energy Engineering, Shanghai University,
Shanghai 200072, People\rq{}s Republic of China
}


\author{Jianping Meng (ÃϽ£Æ½)}

\affiliation{James Weir Fluids Lab, Department of Mechanical and Aerospace Engineering, University of Strathclyde, Glasgow G1 1XJ }

\date{\today}

\begin{abstract}
In this paper, the scaling property of the inverse energy cascade and  forward enstrophy cascade of the vorticity filed  $\omega(x,y)$
in two-dimensional (2D) turbulence is analyzed. This is accomplished by applying a Hilbert-based technique, namely Hilbert-Huang Transform, to a vorticity  field obtained from a $8192^2$ grid-points direct numerical simulation
of the 2D turbulence with a forcing scale $k_f=100$ and an Ekman friction.
The 
measured joint probability density function  $p(C,k)$ of mode $C_i(x)$ of the vorticity $\omega$ and instantaneous wavenumber $k(x)$ is separated by the forcing scale $k_f$ into two parts, which corresponding to the  inverse energy cascade and the forward enstrophy cascade. It is found that all conditional pdf $p(C\vert k)$  at given wavenumber $k$
   has an 
exponential tail. In the inverse energy cascade, the shape of $p(C\vert k)$ 
does collapse with each other, indicating a nonintermittent cascade.
The measured  scaling 
exponent $\zeta_{\omega}^I(q)$ is  linear with the statistical order $q$, i.e., $\zeta_{\omega}^I(q)=-q/3$, confirming the nonintermittent cascade process. 
   In the forward 
enstrophy cascade, the core part of $p(C\vert k)$ is changing with  wavenumber $k$, indicating an intermittent forward cascade. 
   The measured  scaling exponent $\zeta_{\omega}^F(q)$  is nonlinear with $q$ and can be described very well by a log-Poisson fitting: $\zeta_{\omega}^F(q)=\frac{1}{3}q+0.45\left(  1-0.43^{q}\right)$.  
However, the extracted vorticity scaling exponents $\zeta_{\omega}(q)$ for both inverse energy cascade and forward enstrophy cascade are not consistent with Kraichnan\rq{}s theory prediction. New theory for the vorticity field in 2D turbulence is required to interpret the observed scaling behavior. 
  \end{abstract}

\maketitle
\end{CJK*}

\section{Introduction}

Two-dimensional (2D) turbulence is an ideal model for several  turbulent phenomena, such as the
first approximation to the large-scale motion in atmosphere and oceans.
\cite{Kraichnan1980RepProgPhys,Tabeling2002PhysRep,
Kellay2002RepProgPhys,Boffetta2012ARFM,
Bouchet2012PhysRep}
The 2D turbulence and relative problems have 
attracted a lot of attentions in recent years. \cite{Irion1999Science,Falkovich1994PRE,Chen2003PRL,Chen2006PRL,Boffetta2010PRE,Alexakis2006PLA,
Xia2011NatPhys,Xia2008PRL,Tran2010NaturePhys,Kelley2011PoF,Merrifield2010PRL,Celani2010PRL,Khurana2012PoF} Several review papers have been 
devoted to this topic in a detail, for example, papers by \citet{Tabeling2002PhysRep,
Kellay2002RepProgPhys,Heijst2009ARFM,Boffetta2012ARFM,
Bouchet2012PhysRep}, to quote a few.
 \red{The 2D Ekman-Navier-Stokes equation is written in term of a single 
scalar vorticity  field $\omega=\nabla  \times \mathbf{u}$  as, i.e.,
\begin{equation}
\partial_t \omega+\mathbf{u}\bullet \nabla \omega =\nu \bigtriangledown^2 \omega-\alpha \omega +f_{\omega}  \label{eq:NSvorticity}
\end{equation}
in which    $\nu$ is the fluid viscosity, $\alpha$ is the Ekman friction and $f_{\omega}$ is an external source of energy inputing into the whole system. \cite{Boffetta2002PRE,Boffetta2007JFM}}
    Specifically for the small scale motions, it is believed that 
there exists a dual-cascade, i.e., a forward enstrophy cascade, in which the enstrophy (square of vorticity $\omega^2$) is transfered from large  to small scales,  and an inverse energy 
cascade, in which the energy is transfered from small to large scales. \cite{Kraichnan1967PoF}  A two-power-law behavior is thus expected to describe this dual cascade, i.e.,
\begin{equation}
E_u(k)= \left\{
\begin{array}{lll}
&C\left(\epsilon_{\alpha}\right)^{2/3}k^{-5/3}, &\textrm{ when $k_{\alpha}\ll k \ll k_f$ for inverse  energy cascade}\\
&C\rq{} \left(\eta_{\nu}\right)^{2/3} k^{-3}, &\textrm{ when $k_{f} \ll  k\ll k_{\nu}$ for \red{forward enstrophy cascade}}
\end{array}
\right.
\label{eq:velocity}
\end{equation}
in which  $E_u(k)$ is Fourier power spectrum of the velocity, $\epsilon_{\alpha}$ is the energy dissipation by the Ekman friction, $\eta_{\nu}$ is the enstrophy dissipation by the viscosity, $k_f$  is the forcing scale, in which the energy and enstrophy is injected into the system, and $k_{\alpha}$ is the characteristic friction scale, $k_{\nu}$ is the viscosity scale.
One can relate the vorticity statistics with the velocity one by using $E_{\omega}(k)\sim k^2 E_{u}(k) $. Therefore, a dual power-law behavior is also expected for the vorticity field, i.e.,
\begin{equation}
E_\omega(k)\sim  k^2E_{u}(k) \sim \left\{
\begin{array}{lll}
&k^{1/3}, &\textrm{ when $k_{\alpha}\ll k \ll k_f$ for \red{inverse energy cascade}}\\
& k^{-1}, &\textrm{ when $k_{f} \ll  k\ll k_{\nu}$ for forward enstrophy cascade}
\end{array}
\right.
\label{eq:vorticity}
\end{equation}   It is found experimentally that  the pdf of the velocity increment $\Delta_{\ell} u$ is Gaussian when the separation scale $\ell$ lies in the inverse energy cascade, indicating nonintermittent behavior on these scales. \cite{Tabeling2002PhysRep,
Kellay2002RepProgPhys,Heijst2009ARFM,Boffetta2012ARFM,
Bouchet2012PhysRep}
Note that the classical structure-function (SF) analysis fails when the slope $\beta$ of the Fourier 
power spectrum is out of the range $1<\beta< 3$, in which $E(k)\sim k^{-\beta}$. \cite{Frisch1995,Huang2010PRE} This 
unfortunately is the case of the forward enstrophy cascade in the 2D turbulence.
\cite{Boffetta2012ARFM,Biferale2003PoF} Therefore, the intermittent property of the forward enstrophy cascade 
can not be verified directly by using the SF analysis. 
\cite{Boffetta2002PRE,Biferale2003PoF}
\citet{Kellay1998PRL} performed an experimental measurement of  the velocity and vorticity field of the  2D soap turbulence.  They found that the  \red{Fourier power spectrum of the} velocity shows a $-3$ power-law for the forward enstrophy cascade, \red{which agrees very well with the theory}. However, the corresponding \red{Fourier power spectrum of the vorticity field for the} forward enstrophy cascade demonstrates a $-2$ power-law, which is contradicted with the theoretical prediction, see Eq.\,\eqref{eq:vorticity}. 
For the velocity measurement, 
 \citet{Paret1999PRL} also observed $-3$  scaling  for the forward enstrophy cascade. Moreover, the measured pdf of vorticity increment $\Delta_{\ell} \omega$  is not significant deviation 
from the Gaussian distribution, i.e., a nonintermittent forward enstrophy cascade.  On the 
contrary,  \citet{Nam2000PRL} found  that if an Ekman friction coefficient $\alpha$ is presented, the forward enstrophy cascade  is then intermittent. \cite{Bernard2000EPL}  \citet{Boffetta2002PRE}  argued that if a passive scalar $\theta$  is governed  by the same equation as the vorticity field and if it is also advected by  the same velocity field,  it then can be taken as a surrogate of the vorticity  $\omega$ for the small scale statistics. 
 They found that the passive scalar $\theta$ is indeed  intermittent. Moreover, they found that the fitting scaling exponent for the forward enstrophy  cascade is dependent 
 on the Enkman viscosity $\alpha$.
 Later,  \citet{Tsang2005PRE} studied  the intermittency of the forward enstrophy  cascade 
 regime with a linear drag. The relative scaling 
 exponent ($\zeta(2q)/\zeta(2)$) provided by the vorticity SF 
 confirms 
 that the forward  enstrophy cascade is  intermittent for the considered 
 statistical order $0\le q \le 2$.  Note that the classical 
 SF approach is employed in their studies.   \citet{Biferale2003PoF} 
 proposed  an inverse velocity statistics and applied in 2D turbulence. They found 
 that the velocity fluctuation can not be simply described by one single exponent, 
 indicating an intermittent forward  cascade.  \citet{Boffetta2007JFM} reported that 
 the fitting scaling exponent of \red{the Fourier power spectrum} for the forward  
 cascade might also depend
 on the  viscosity $\nu$.
Recently, \citet{Falkovich2011PRE} 
derived analytically the probability density function (pdf) for strong vorticity 
fluctuations (resp. the tail of the pdf) in the forward enstrophy cascade. They found that the  
over $R$ coarse-grained vorticity $\overline{\omega}(R)$  has a universal asymptotic 
exponential tail and is thus
 self-similar without intermittency (resp.scaling exponent is linear with $q$) at least for high-order statistics.  Generally 
 speaking, Kraichnan\rq{}s theory of 2D turbulence is partially confirmed by the 
 experiments and numerical simulation for the velocity field.
 \cite{Falkovich2006PhysToday} However, as mentioned above, the statistics of the 
 vorticity field seems to disagree with the theoretical prediction.

 In this paper, we apply a Hilbert-based technique to  the 
 vorticity $\omega(x,y)$ field  obtained from a high resolution direct numerical simulation (DNS). 
 A dual-cascade behavior is observed respectively with a nearly one decade inverse energy cascade and forward enstrophy cascade. For the inverse energy cascade,  the measured  vorticity pdf   does 
 collapse with each other, implying a nonintermittent  cascade 
 process as expected. \cite{Falkovich2011PRE,Boffetta2012ARFM,Bouchet2012PhysRep} The corresponding  measured scaling exponent $\zeta_{\omega}^I(q)$ is linear with $q$, i.e., $\zeta_{\omega}^I=q/3$.
 For the  forward enstrophy cascade,  the measured
 vorticity pdf  possesses an exponential tail, which is consistent with the  findings in Ref. \cite{Falkovich2011PRE}.  \red{However, they can not collapse with each other, indicating an intermittent forward enstrophy cascade}.  The measured scaling exponent $\zeta_{\omega}^F(q)$ is nonlinear with a small $q$ and   asymptotic to a linear relation for large $q$. A log-Poisson-like formula is proposed to describe the measured scaling exponent, i.e.,  $\zeta^F_{\omega}=q/3+0.45(1-0.43^q)$ for  the forward enstrophy cascade. \red{Note that for the vorticity field, the measured scaling exponents disagree with the theoretical prediction by \citet{Kraichnan1967PoF} even one takes the logarithmic correction into account. New theory about the vorticity field  is required to interpret our findings in this work.}

\section{Hilbert-Huang Transform}

\begin{figure*}[htb]
\centering
 \includegraphics[width=0.65\linewidth,clip]{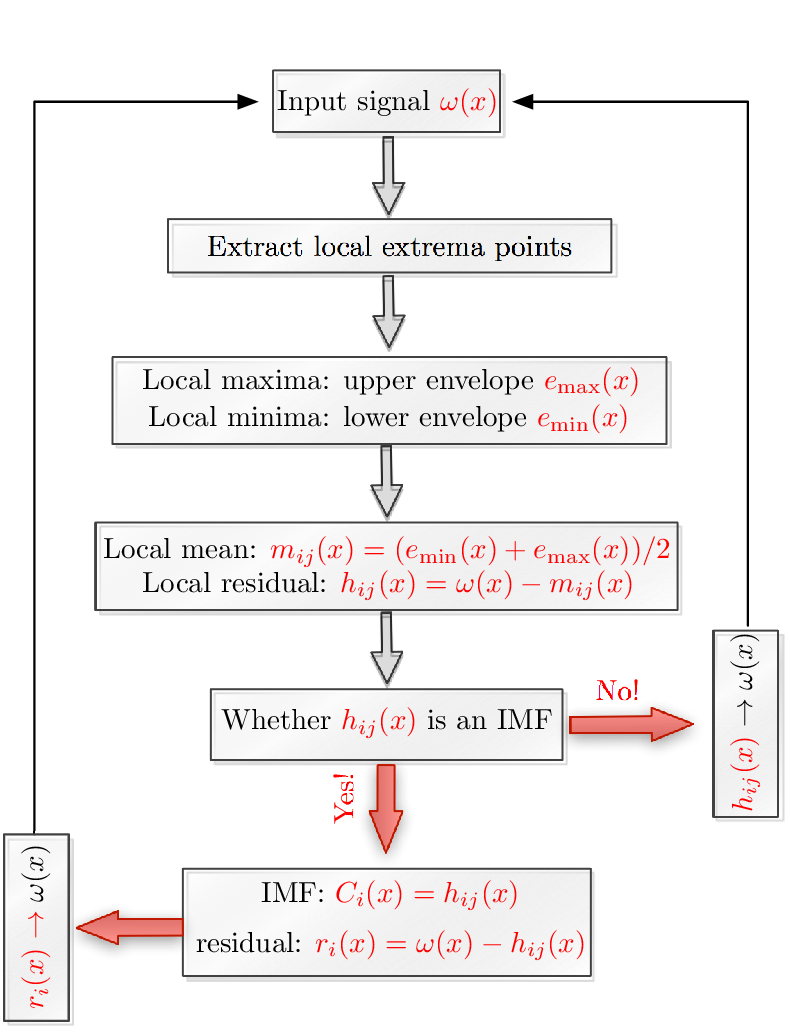}
  \caption{(Color online) A flowchart of the Empirical Mode Decomposition algorithm to show how to decompose a given vortex filed $\omega(x,y)$ at certain $y$ into a sum of Intrinsic Mode Functions $C_i(x)$.  }\label{fig:EMD}
\end{figure*}

\subsection{Empirical Mode Decomposition}

The method we used in this work is  the so-called arbitrary-order Hilbert spectral 
analysis. \cite{Huang2008EPL,Huang2011PRE} It is an extended version of the
Hilbert-Huang Transform (HHT). \cite{Huang1998EMD,Huang1999EMD}
The Hilbert method contains two steps. In the first step,  a data-driven algorithm, 
namely Empirical Mode Decomposition (EMD),  is designed to decompose a given signal, 
e.g., $\omega(x)$, into  a sum of Intrinsic Mode Functions (IMFs) $C_i(x)$
without \textit{a priori} basis.  \cite{Huang1998EMD,Flandrin2004EMDa}
The IMF is an approximation of the 
 mono-component signal, which has to satisfy the following two conditions:  (\romannumeral1)
 the difference between the number of local extrema and the number 
 of zero-crossings must be zero or one; 
 (\romannumeral2) the running mean value 
 of the envelope defined by the local maxima and the envelope 
 defined by the local minima is zero. 
 \citep{Huang1998EMD,Huang1999EMD,Rilling2003EMD} The extracted IMF mode 
 possesses a well-behave Hilbert spectrum with a physical meaningful instantaneous 
 frequency in time domain (resp. wavenumber in space domain).\cite{Huang1998EMD,Huang1999EMD} Figure \ref{fig:EMD} shows a 
 flowchart of the EMD algorithm to demonstrate how to decompose a given vortex signal $\omega(x)$ into a sum of IMF modes, i.e.,
\begin{equation}
\omega(x)=\sum_{i=1}^{N} C_i(x)+r_N(x)
\end{equation}
in which $r_N(x)$ is residual. 
There exist several criteria to stop the sifting process and to determine whether an IMF mode is retrieved.\cite{Huang1998EMD,Huang1999EMD,Rilling2003EMD} For example, \citet{Huang1998EMD} has proposed a Cauchy like criteria computed from two consecutive sifting, i.e.,
\begin{equation}
SD=\frac{\int \left[h_{i(j-1)}(x)-h_{ij}(x)\right]^2\upd x} {\int \left[h_{i(j-1)}(x)\right]^2 \upd x}
\end{equation} 
\red{in which $h_{ij}(x)$ is the residual by removing the running mean $m_{ij}(x)$ constructed by using upper $e_{\max}(x)$ and lower envelopes $e_{min}(x)$. Here $e_{\max}(x)$ (resp. $e_{\min}(x)$) is the upper envelope constructed by using the local maxima points (resp. minima points). }
A typical value can be set between $0.2$ and $0.3$ to provide a physical meaningful 
IMF mode. Another widely used stopping criteria is proposed by 
\citet{Rilling2003EMD}. They introduced an amplitude function $a(x)=(e_{\max}(x)-
e_{\min}(x))/2$ and an evaluation function $\sigma(x)=m_{ij}(x)/a(x)$, respectively. 
The sifting procedure is iterated until $\sigma(x)<\theta_1$ for some prescribed 
fraction $(1-\alpha)$ of the total data, while $\sigma(x)<\theta_2$ for the rest 
fraction. Typical values are $\alpha\simeq 0.05$, $\theta_1\simeq0.05$ and 
$\theta_2\simeq 10\theta_1$.\cite{Rilling2003EMD} 
A maximum iteration number, i.e., $100$, could also be used  to stop the sifting. 
In our practice, any one of the above three criteria is satisfied, then one IMF mode is retrieved.

\subsection{Hilbert Spectral Analysis}  
   
After retrieving the IMF modes,  the  Hilbert spectral analysis is  applied to each mode to obtain the 
time-frequency information. \cite{Huang1998EMD,Huang1999EMD,Huang2009PHD} 
The Hilbert transform is defined as, i.e.,
\begin{equation}
\overline{C}_i(x)=\frac{1}{\pi}P\int \frac{C_i(x')}{x-x'}\upd x'\label{eq:Hilbert}
\end{equation}
in which $P$ means the Cauchy principle  value.\cite{Cohen1995,Flandrin1998}
A so-called analytical signal is then written as, i.e., 
\begin{equation}
C_i^A(x)=C_i(x)+j\overline{C}_i(x)
\end{equation}
in which $j=\sqrt{-1}$.
An instantaneous wavenumber is then defined as, i.e.,
\begin{equation}
k_i(x)={\frac{1}{2\pi}} \frac{\upd}{\upd x} \arctan \left(\frac{\overline{C}_i(x)}{C_i(x)}\right)\label{eq:wavenumber}
\end{equation}
Note that Eq.\,\eqref{eq:Hilbert} is a singularity transform. The differential operation is also used to define the instantaneous wavenumber, see Eq.\,\eqref{eq:wavenumber}.   Therefore, the Hilbert method possesses a very local ability in the wavenumber domain.  
\red{The final representation of the original data $\omega(x)$ can be written as, i.e.,
\begin{equation}
\omega(x)=\sum_{i=1}^N\mathcal{A}_i(x)\exp \left(\int_{-\infty}^x j k_i(x\rq{}) \upd x\rq{} \right)
\end{equation}
in which $\mathcal{A}_i(x)=\sqrt{\overline{C}_i(x)^2+C_i(x)^2} $  is the modulus of $C_i^A(x)$. \cite{Huang1998EMD,Huang2011PRE} Comparison with the Fourier analysis, the above representation can be considered as a local Fourier expansion, in which the amplitude $\mathcal{A}$ and wavenumber $k$ can be varied with $x$, namely 
amplitude- and frequency-modulation.\cite{Huang1998EMD}
}

\subsection{Arbitrary Order Hilbert Spectral Analysis}

After obtaining the IMF modes $C_i(x)$ and the corresponding instantaneous 
wavenumber $k_i(x)$, one can construct a set of pair $[C_i(x), k_i(x)]$.
 A  $k$-conditional $q$th-order statistical moment  is then 
defined as, i.e.,
\begin{equation}
\mathcal{L}_q(k)=\langle \sum_{i=1}^{N} \left[ C_i (x)\vert_{k_i(x)=k} \right]^q \rangle_{x}\label{eq:LQ1}
\end{equation}
in which $\langle \cdots \rangle$ is ensemble average \red{over all $i$ and $x$}. \cite{Huang2011PRE,Huang2013PRE}
 $\mathcal{L}_q(k)$ could be defined by another equivalent way as described below.
 One can extract a joint probability density function (pdf)  i.e., $p(C,k)$, from the IMF mode $C_i(x)$ and the corresponding wavenumber $k_i(x)$. 
Taking a marginal integration,  Eq.\,\eqref{eq:LQ1}  is then rewritten as, i.e.,
 \begin{equation}
 \mathcal{L}_q(k)=\frac{\int p(C,k)\vert C\vert^q \red{\upd C}}{\int p(C,k) \red{\upd C}}
 \end{equation}
 For a scaling process, one expects a power-law behavior, i.e.,
 \begin{equation}
 \mathcal{L}_q(k)\sim k^{-\zeta(q)}
 \end{equation}
 in which $\zeta(q)$ is comparable with the scaling exponents provided by the 
 classical structure function.\cite{Huang2011PRE}  
 
\red{ The Hilbert-based methodology has been verified by using a synthesized fractional Brownian motion data for mono-fractal process and a synthesized multifractal random walk with a lognormal statistics for multifractal
 process. \cite{Huang2011PRE} It also has been applied successfully to turbulent velocity,\cite{Huang2008EPL} passive scalar, \cite{Huang2010PRE} Lagrangian 
turbulence,\cite{Huang2013PRE} etc., to characterize the intermittent nature of those processes. 
Our experience is that the SF analysis works when  $1<\beta<3$ without energetic structures.\cite{Huang2010PRE}  Here $\beta$ is the 
scaling exponent of Fourier power spectrum, i.e., $E(k)\sim k^{-\beta}$. If $\beta$ is out 
of this range, then the Hilbert methodology should be applied to extract scaling exponents for high-order $q$. Moreover, due to the influence of energetic structures, the SFs may fail even when $\beta=2$. This has been found to be  the case of the three-dimensional Lagrangian turbulence.\cite{Huang2013PRE} We will show below that this is also the case of 2D turbulence, see more discussion in Sec.\,\ref{sec:SF}.}
For more detail about the methodology  we refer the reader to the Refs. 
 \onlinecite{Huang1998EMD,Huang2008EPL,Huang2009PHD,Huang2011PRE,Huang2013PRE}.

\section{Numerical Data and scaling of high-order statistics}

\begin{figure*}[!htb]
\centering
 \includegraphics[width=0.65\linewidth,clip]{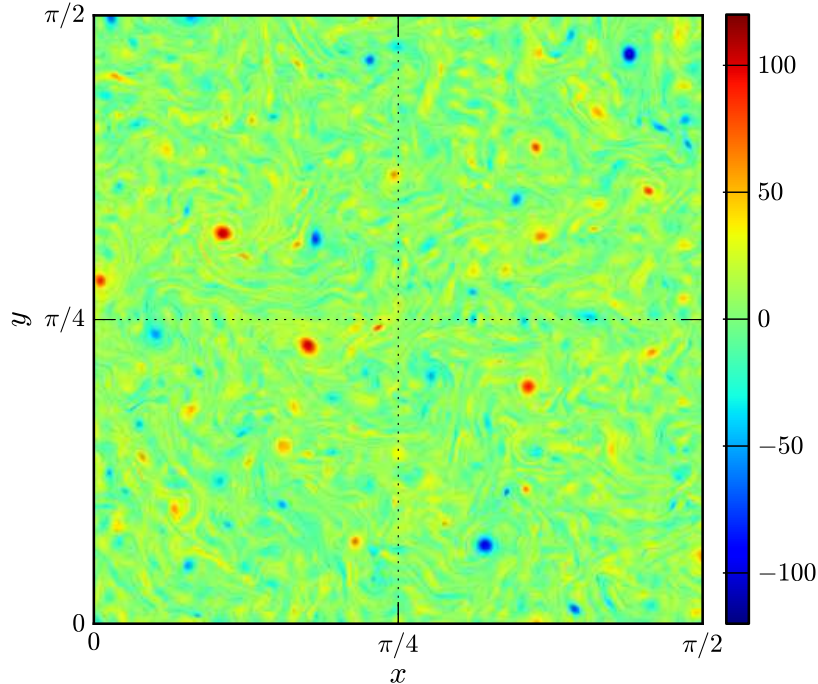}
  \caption{(Color online) A snapshot of the vorticity field $\omega(x,y)$ of the two-dimensional turbulence on the range $0\le x,y\le \pi/2$ from a high resolution direct numerical simulation with $8192^2$ grid points. High intensity vorticity events are discrete distributed in space with a typical wavenumber $k\simeq k_f=100$ (resp. around $80$ grid points).}\label{fig:snapshot}
\end{figure*}

 \begin{figure*}[!htb]
\centering
 \includegraphics[width=0.95\linewidth,clip]{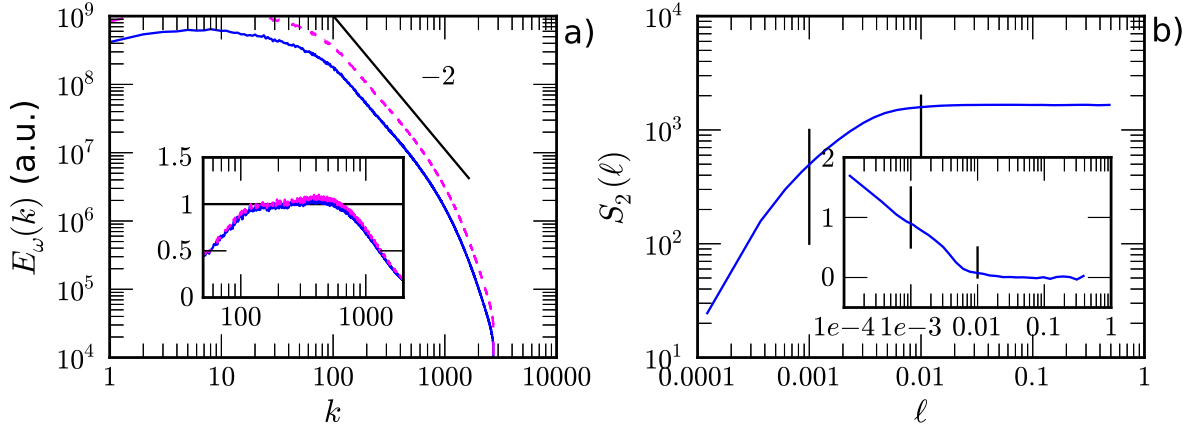}
  \caption{(Color online) a)  Measured Fourier power spectrum $E_{\omega}(k)$ for vorticity field $\omega(x,y)$, \red{in which a logarithmic correction $E_{\omega}(k)\ln(k/k_{\min})^{-1/3}$ is shown as a dashed line. For display clarity, the measured $E_{\omega}(k)\ln(k/k_{\min})^{-1/3}$  has been vertical shifted by multiplying a factor 5.} A nearly one decade power-law behavior is found on the wavenumber range $ 100\le k \le 1000 $, corresponding to $0.001\le \ell \le 0.01$ with a scaling exponent $\beta=1.96\pm0.02$  \red{for $E_{\omega}(k)$ and $\beta=2.02\pm0.02$ for $E_{\omega}(k)\ln(k/k_{\min})^{-1/3}$, respectively}. The inset shows the compensated spectrum  with fitted scaling exponents. This scaling range corresponds to the forward  enstrophy  cascade. b) Measured second-order structure-function $S_{\omega}(2,\ell)$. No power-law behavior is observed as expected  on the range $0.001\le \ell \le 0.01$. The inset shows the local slope $\zeta_{\omega}(2,\ell)=\upd \log_{10} S_{\omega}(2,\ell)/\upd \log_{10} \ell$ to confirming the lacking of the power-law behavior.}\label{fig:psd}
\end{figure*}

 \begin{figure*}[!htb]
\centering
 \includegraphics[width=0.65\linewidth,clip]{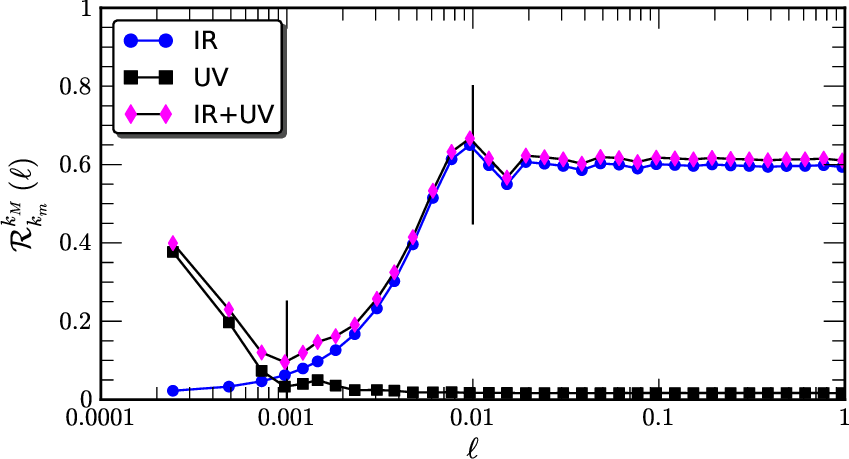}
  \caption{(Color online)   Measured relative contribution function $\mathcal{R}^{k_M}_{k_m}(\ell)$ from different Fourier modes. Low wave number part (IR) $[0,100]$ ($\ocircle$), high wavenumber part (UV) $[1000,+\infty]$ ($\square$). The expected power-law range is illustrated by a vertical solid line. Note that the expected power-law  behavior is strong influenced by the low wavenumber part, known as IR effect. }\label{fig:contribution}
\end{figure*}

\subsection{Direct Numerical Simulation of 2D Turbulence}

\red{The DNS data we used in this study is provided by Professor G. Boffetta. We recall briefly several key parameters of this simulation.}
Numerical integration of Eq.\,\eqref{eq:NSvorticity} is  performed by a pseudo-spectral, fully dealiased on a doubly periodic square domain of size $L=2\pi$ at resolution $N^2=8192^2$ grid points.
\cite{Boffetta2007JFM}  The main parameters  are  respectively $\nu=2\times10^{-6}$, $\alpha=0.025$ and $k_f=100$, in which the energy is 
inputed into the system. 
The velocity field $ \mathbf{u}=\nabla  \times \Phi$ is then obtained by solving a 
Poisson problem $\nabla ^2 \Phi=-\omega$, in which $\Phi$ is a stream function. 
\red{Totally, we have five snapshots with $8192^2\times 5=3.36\times10^8$ data points. In the following, the analysis is done along the $x$-direction. This provides $8192\times 5=40960$ realizations for each statistics. The ensemble average is then averaged from all these realizations.}
Figure \ref{fig:snapshot} shows a portion of one snapshot of the vorticity field. Note 
that high intensity events are discrete distributed in physical space with a typical 
wavenumber $k\simeq k_f=100$ (resp. $\sim 80$ grid points).  
More detail of this database can be found in 
Ref.\,\onlinecite{Boffetta2007JFM}.

\subsection{Fourier Power Spectrum  and Second-Order Structure-Function}\label{sec:SF}

Figure \ref{fig:psd}\,a) shows the measured Fourier power spectrum $E_{\omega}(k)$ (solid line) of the vorticity field, \red{in which the logarithmic correction $E_{\omega}(k)\ln(k/k_{\min})^{-1/3}$ is shown as a dashed line}.  A power-law behavior for the forward enstrophy cascade is observed on the range $100\le k \le 1000$, i.e., $E_{\omega}(k)\sim k^{-\beta}$, with a scaling exponent $\beta=1.96\pm0.02$.
The measured $\beta$ is consisted with the one reported by   \citet{Kellay1998PRL}.
  The observed scaling range corresponds to a spatial scale range $0.001\le \ell \le 0.01$. \red{ The logarithmic correction provides a  scaling exponent $\beta=2.02\pm0.02$ on the same scaling range, showing a weak correction of the power-law behavior.\cite{Pasquero2002PRE} The inset shows the corresponding compensated curves to emphasize the observed power-law behavior.}
We therefore expect a power-law behavior on the range $0.001\le \ell\le0.01$ for the second-order SFs, i.e., 
\begin{equation}
S_{\omega}(2,\ell)=\langle \vert\Delta_{\ell} \omega \vert^2\rangle\sim \ell^{\beta-1}
\end{equation}
in which $\Delta_{\ell} \omega=\omega(x+\ell)-\omega(x)$  is vorticity increment, 
and $\beta$ is the scaling exponent from $E_{\omega}(k)\sim k^{-\beta}$. 
Figure \ref{fig:psd}\,b) shows the measured second-order SF, in which the forward enstrophy cascade is illustrated by a vertical solid line. Visually,  no  power-law 
behavior is observed for the measured $S_{\omega}(2,\ell)$ for the forward enstrophy cascade. To emphasize this point, the local slope, i.e., $\zeta_{\omega}(2,\ell)=\upd \log_{10} S_{\omega}(2,\ell)/\upd \log_{10} \ell$, is shown in the inset.  There is no plateau observed on the range of the forward enstrophy cascade, showing the failure of the SFs to capture the scale invariance of the two-dimensional vorticity field.

To understand more about  the second-order SFs $S_{\omega}(2,\ell)$, one can relate it with the Fourier power spectrum by using Wiener-Khinchin theorem,\cite{Frisch1995,Huang2010PRE}  i.e.,
\begin{equation}
S_{\omega}(2,\ell)=\int_{0}^{+\infty}E_{\omega}(k)(1-\cos(2\pi k\ell))\upd k
\end{equation}
in which $E_{\omega}(k)$ is the corresponding Fourier power spectrum. Note that an
integral constant is neglected since it does not change the conclusion in this paper. 
The above equation implies that the SFs contains contribution from almost all Fourier modes $k$. The expected power-law behavior might be influenced by both large-scale (resp. low wavenumber, known as  infrared effect, IR) and small-scale (resp. high wavenumber, known as ultraviolet effect, UV) motions.
A partial cumulative function is therefore introduced to characterize a relative contribution from Fourier modes band $[k_m, k_M]$, i.e.,
\begin{equation}
\mathcal{R}_{k_m}^{k_M}(\ell)=\frac{\int_{k_m}^{k_M}E_{\omega}(k\rq{})(1-\cos(2\pi k\rq{}\ell))\upd 
k\rq{}}{\int_{0}^{+\infty}E_{\omega}(k)(1-\cos(2\pi k\ell))\upd k}
\end{equation}
We are particular  concerned by the Fourier modes below the forcing scale,i.e., $[0,100]$ (resp. IR)  and by the modes above the power-law range, i.e., $[1000,+\infty] $ (resp. UV).
Figure \ref{fig:contribution} shows the measured $\mathcal{R}_{m}^{M}(\ell)$ for IR 
($\ocircle$) and UV ($\square$). It indicates that the SFs is 
strongly influenced by the large-scale motions (resp. IR) as high as up to $65\%$. 
 \red{The expected power-law behavior is then destroyed.}\cite{Davidson2005PRL,Huang2010PRE}

We provide some comments on the above analysis here.  The Wiener-Khinchin 
theorem is only exactly valid for the linear and stationary processes.
Turbulent signals in 3D or 2D are typical nonlinear and nonstationary ones. Therefore, the above argument holds approximately  here. However, this does not change the conclusion of this paper.  \red{Another comment is for the observed high intensity vorticity event, see Fig.\,\ref{fig:snapshot}. For the high-order SFs, they might be also  influenced by those events since they usually manifest themselves at the pdf tail of vorticity increments. This has been observed for the Lagrangian turbulence, in which the high intensity event is known as \lq{}vortex trapping\rq{} process.\cite{Toschi2005JOT,Huang2013PRE} }

 \subsection{Generalization Scaling for High-Order Statistics}

\red{Assuming that we have SFs scaling for both the inverse and forward cascades. The corresponding SFs and their scaling exponents without intermittent corrections are
\begin{subequations}\label{eq:HQ}
\begin{equation}
S_{u}(q,\ell)\sim \left\{
\begin{array}{lll}
&\ell^{q/3}, &\textrm{\red{for inverse energy cascade}}\\
& \ell^{q}, &\textrm{for forward enstrophy cascade}
\end{array}
\right.
\end{equation}
for the velocity field,\cite{Falkovich1994PRE}
and 
\begin{equation}
S_{\omega}(q,\ell)\sim \ell^{-q}S_{u}(q,\ell)\sim \left\{
\begin{array}{lll}
&\ell^{-2q/3}, &\textrm{\red{for inverse energy cascade}}\\
& \ell^{0}, &\textrm{for forward enstrophy cascade}
\end{array}
\right.
\end{equation}
for the vorticity field. It implies that the forward enstrophy cascade represented by the vorticity 
field is independent on the separation scale $\ell$. In the frame of Hilbert, we expect the following scaling behavior for the vorticity field, i.e.,
\begin{equation}
\mathcal{L}_{\omega}(q,k)\sim  \left\{
\begin{array}{lll}
&k^{2q/3}, &\textrm{\red{  for inverse energy cascade}}\\
& k^{0}, &\textrm{  for forward enstrophy cascade}
\end{array}
\right.\label{eq:Hilbertscaling}
\end{equation}
\end{subequations}
}
We will then test above relation by using the Hilbert method as we described above.

\section{Hilbert Results and Discussion}

\subsection{Hilbert Statistics}
 
 \begin{figure*}[!htb]
\centering
 \includegraphics[width=0.95\linewidth,clip]{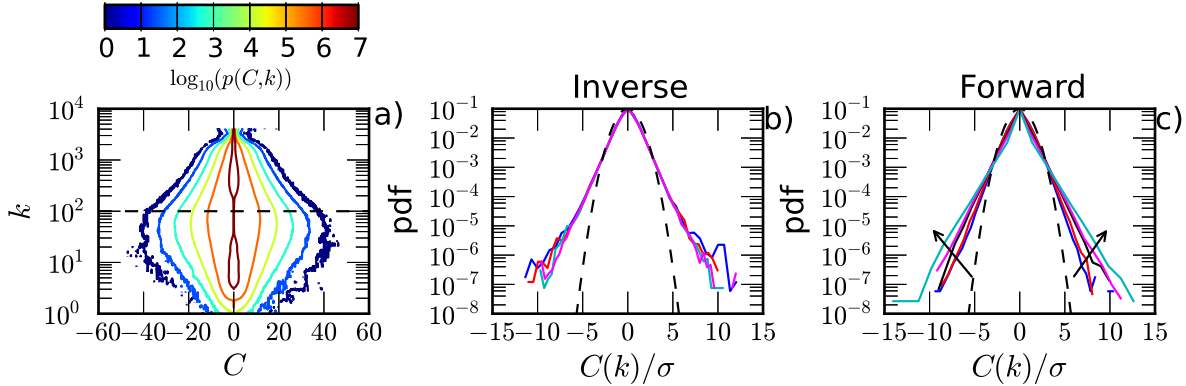}
  \caption{(Color online) a)  Contour plot of measured  conditional histogram $p(C,k)$ (resp. joint probability density function  (pdf)), in which the force scale is illustrated by a dashed line. b) Measured pdf $p(C)$ \red{on the range  $3<k<20$ (resp. $k=4,6,16$ and $20$) for the  inverse energy cascade }. \red{c) Measured pdf $p(C)$ on the range   $200<k<2000$ (resp. $k=250,400,800,1000$ and $1600$) for the forward enstrophy cascade.}   For comparison, the normal distribution is illustrated by a dashed line.  }\label{fig:pdf}
\end{figure*}

 \begin{figure*}[!htb]
\centering
 \includegraphics[width=0.65\linewidth,clip]{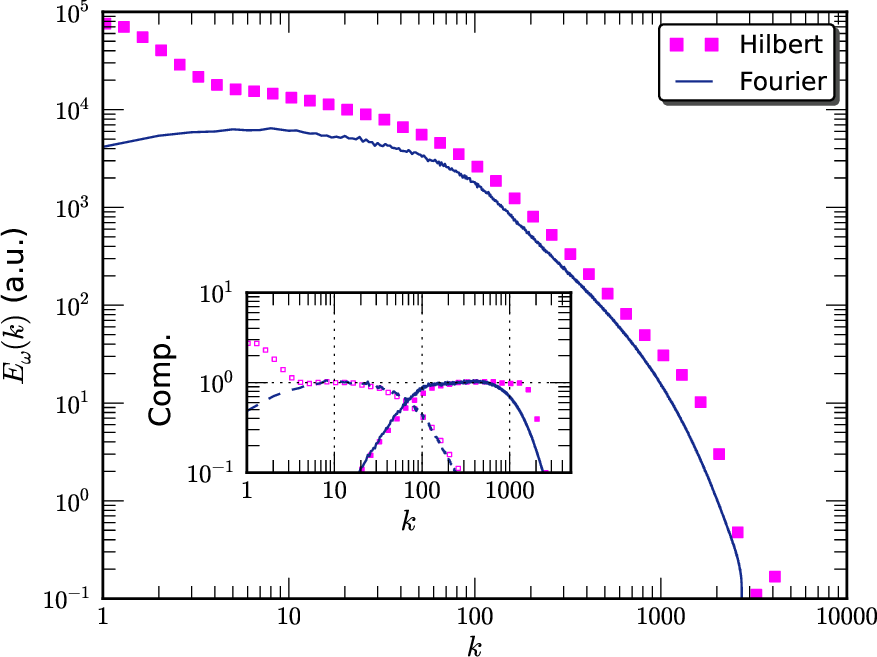}
  \caption{(Color online) \red{Comparison  of the measured Hilbert energy spectrum $\mathcal{L}_2(k)/k$ ($\square$) and the Fourier energy spectrum $E_{\omega}(k)$ (solid line). For display clarity, the curves have been vertical shifted.  Inset shows the corresponding compensated spectra for both inverse energy cascade and forward enstrophy cascade by using  fitting scaling exponents respectively on the range $3<k<20$ and $100<k<1000$.  }}\label{fig:spectrum}
\end{figure*}

\begin{figure*}[!htb]
\centering
 \includegraphics[width=0.95\linewidth,clip]{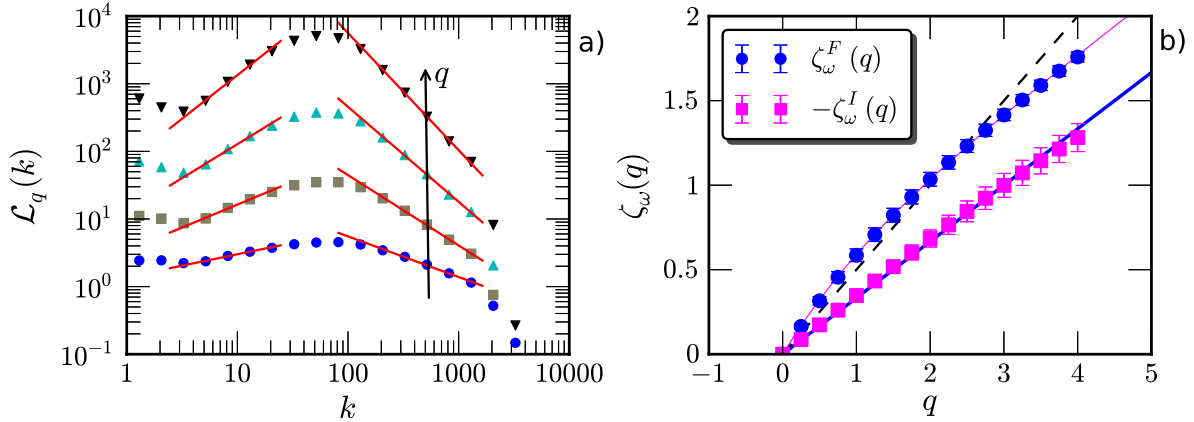}
  \caption{(Color online) 
  a) Measured $q$th-order Hilbert spectra $\mathcal{L}_q(k)$ for $q=1,2,3,4$. power-law behavior is observed on the range $200\le k \le 2000$ for the forward enstrophy cascade and on the range $3\le k \le 20$ for the inverse energy cascade, respectively. The scaling exponents $\zeta_{\omega}(q)$ are then estimated on this dual power-law ranges. b) Measured scaling exponents $\zeta_{\omega}^F(q)$ ($\ocircle$) and $-\zeta_{\omega}^I(q)$ ($\square$). For comparison, the dashed line is $q/3$ and the solid line is for a log-Poisson fitting $\zeta_{\omega}^F(q)=q/3+0.45(1-0.43^q)$. The errorbar indicates $95\%$ fitting confidence.
    }\label{fig:scaling}
\end{figure*}

\begin{figure*}[!htb]
\centering
 \includegraphics[width=0.65\linewidth,clip]{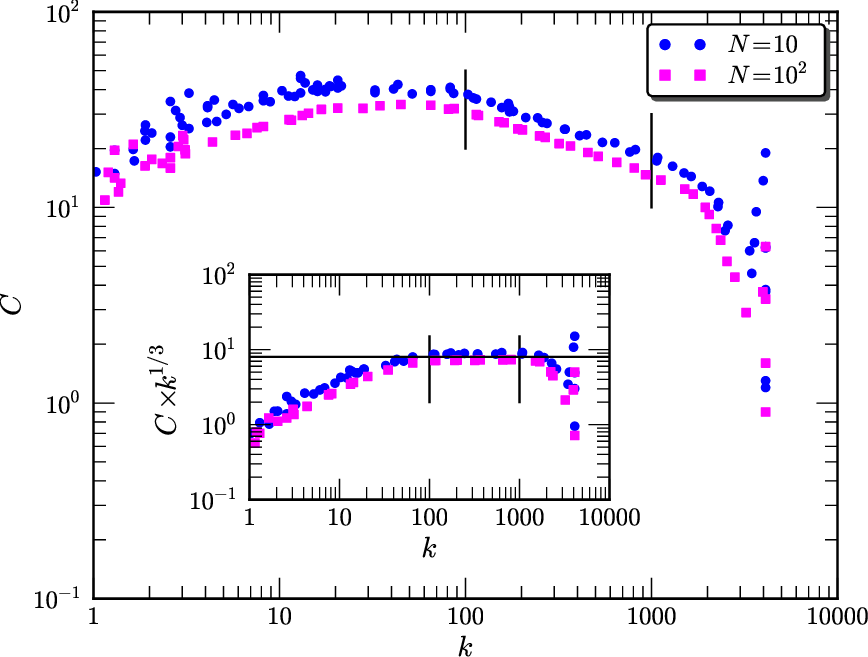}
  \caption{(Color online) 
  Experimental contour line for $N=10$ ($\ocircle$) and $N=100$ ($\square$). power-law behavior is observed on the range $100\le k\le 1000$ with a scaling exponent $\simeq 1/3$. The inset shows the compensated curve by multiplying $k^{1/3}$. The observed power-law indicates an asymptotic scaling exponent $\lim_{q\rightarrow +\infty}\zeta_{\omega}^{F}(q)=q/3$.
    }\label{fig:scalingtrend}
\end{figure*}

The  Hilbert method is  applied to the vorticity field $\omega(x,y)$ along $x$-direction.  The conditioned/joint histogram $p(C,k)$ (resp. probability density function if it is renormalized properly) is extracted \red{from all five snapshots}. 
Figure \ref{fig:pdf}\,a) shows  the contour plot of the measured $p(C,k)$, in 
which the forcing scale $k_f=100$ is illustrated by a horizontal dashed line. For 
display clarity,  we take the logarithm of the measured $p(C,k)$. It is interesting to note that the joint 
pdf is roughly separated by the forcing wavenumber
$k_f=100$ into two regimes.  The first regime is on the range $2\le k \le 20$, 
corresponding to the inverse energy cascade. The another one is on the range $200\le k\le 2000$, corresponding to 
the forward enstrophy cascade. Figure \ref{fig:pdf}\,b) shows the 
pdf $p(C\vert k)$ on the inverse energy cascade (resp. $k=4,6,16$ and $20$ ). For comparison, the normal distribution is demonstrated by a solid line. 
All the measured pdf has an exponential tail.
\red {We note that 
the pdf in the inverse energy cascade does collapse with each other, indicating a nonintermittent cascade.} 
For the  pdf in the forward enstrophy cascade,  they also possess an exponential tail. 
\red{ However, they can not collapse with each other due to different shape of the core part $-5<C(k)/\sigma<5$. }
The exponential  tail  of the vorticity field 
consists with very recently theoretical prediction by \citet{Falkovich2011PRE}. 

\red{Figure \ref{fig:spectrum} shows the comparison of the measured Hilbert energy spectrum $\mathcal{L}_2(k)/k$ ($\square$) and the Fourier power spectrum $E_{\omega}(k)$ (solid line). For display clarity, the curve has been vertical shifted. Note that both methods provide the nearly same forward enstrophy cascade  on the range $100<k<1000$ with a scaling exponent close to $\simeq 2$. This scaling exponent agrees very well with the experimental observation by \citet{Kellay1998PRL}. We show in the inset the compensated curve by using the fitted exponent for the forward cascade (closed square for the Hilbert and solid line for Fourier). The observed plateau confirms the power-law behavior as expected.  Furthermore, we have an additional power-law behavior on the range $3<k<20$ for the inverse energy cascade.  The compensated spectra are also shown in the inset (open square for the Hilbert and dashed line for Fourier) to emphasize the observed inverse energy cascade. Both Hilbert and Fourier methodologies identify almost the same dual power-law behavior.}

We now turn to the high-order Hilbert statistics.
The convergence of the statistical moment $\mathcal{L}_q(k)$ has been verified by 
checking the integral kernel $p(C,k)\vert C\vert^q$ at given scales. A quite  good 
convergence has been found for all wavenumber $k$ up to $q=4$  (not 
shown here).  Figure \ref{fig:scaling}\,a) shows the measured $\mathcal{L}_q(k)$  for 
$q=1,2,3$ and $4$ (from bottom to top). A dual power-law behavior is observed as 
expected respectively on the range $3\le k \le 20$ for the inverse energy cascade and $200\le k \le 
2000$ for the forward enstrophy cascade. The scaling exponent $\zeta_{\omega}
(q)$ is then estimated respectively on these two ranges by using a least square fitting algorithm. 
Figure \ref{fig:scaling}\,b) shows the corresponding measured scaling exponent $\zeta_{\omega}(q)$, in which the errorbar is 95\% fitting confidence. 
For the inverse energy cascade, the measured $\zeta_{\omega}^I(q)$ ($\square$) is linear with $q$, i.e., 
$\zeta_{\omega}^I(q)=-q/3$, confirming that there is no intermittent effect in this inverse cascade process.\cite{Boffetta2012ARFM}  However, the observed $-q/3$ scaling does 
not consist with the prediction of Eq.\,\eqref{eq:Hilbertscaling}. This implies that  for the vorticity field
some important mechanisms are ignored in the previously  dimensional arguments of the inverse energy cascade. For example, if one takes the enstrophy dissipation 
$\eta _{\alpha }$ into account and assumes it as important as the Ekman energy dissipation
  $\epsilon _{\alpha }$, one then has the right scaling behavior, i.e., $\protect
  \mathcal {L}_q(k)\sim \left (\epsilon _{\alpha }\eta _{\alpha } \right
  )^{q/6} k^{q/3}$. This corresponds to a scaling exponent $\zeta
  _{\omega }^I(q)=-q/3$ for the inverse energy cascade. However, this naive 
  dimensional argument should be justified for physical evidence and  for more database. 

The measured scaling exponent $\zeta_{\omega}^F(q)$ ($\ocircle$) is also shown in Fig.\,\ref{fig:scaling}\,b). We note that when $q\le 2$ the measured $\zeta_{\omega}^F(q)$  is nonlinear dependent with $q$, indicating an intermittent effect of the vorticity  field. While when $q\ge 2$, it seems to be linear with $q$ with  a slope $\simeq 1/3$. We propose here a log-Poisson-like model for the observed scaling exponent, i.e.,
\begin{equation}
\zeta_{\omega}^F(q)=\gamma q+\mathcal{B}\left(  1-\varphi^{q}\right)\label{eq:logPoissonOmega}
\end{equation}
in which the parameter $\gamma=1/3$, $\mathcal{B}=0.45$ and $\varphi=0.43$ are determined as following.  For large value of $q$ (if the corresponding statistics exists), the
 $q$th-order Hilbert moments $\mathcal{L}_q(k)$ is thus dominated by the 
tail of the pdf $p(C,k)$. Therefore, if the scaling behavior holds, the measured joint pdf $p(C,k)$ should also show a scaling behavior for the contour line. We extract the measured contour line for $p(C,k)$ (in points) with two values $N=10$ ($\ocircle$) and $100$ ($\square$), see Fig.\,\ref{fig:scalingtrend}.  
Power-law behavior is indeed observed as expected on the range $100\le k \le 2000$. The scaling exponent is found to be $\simeq 1/3$. To emphasize the observed $1/3$ scaling, the compensated curves $C(k)k^{1/3}$ are shown as inset of Fig.\,\ref{fig:scalingtrend}. A clear plateau is observed on the expected range  $100\le k \le 2000$. This yields $\gamma=1/3$. The rest of the parameters $\mathcal{B}$ and $\varphi$ are then obtained by using a least square fitting algorithm.

\subsection{Extended Self-Similarity of Structure-Functions}
\begin{figure*}[!htb]
\centering
 \includegraphics[width=0.85\linewidth,clip]{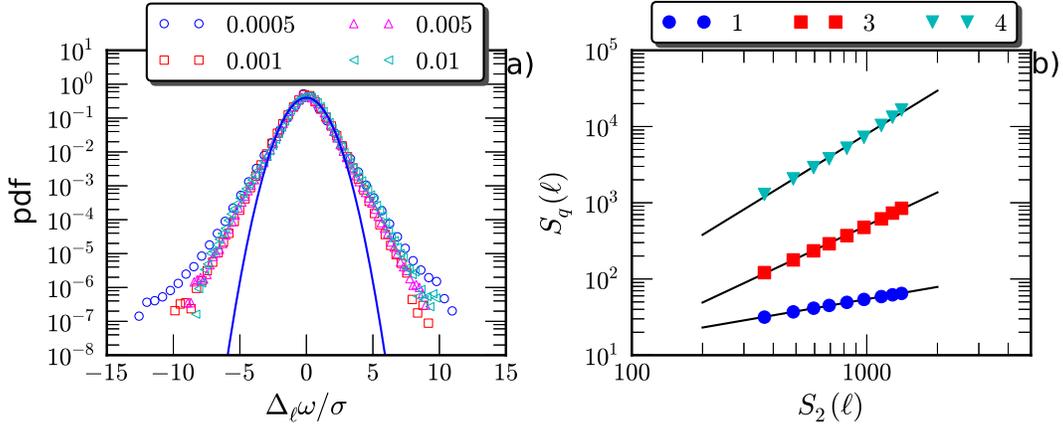}
  \caption{(Color online) \red{ a) Measured pdf for vorticity increments with several separation scales $\ell$ in the range of the forward enstrophy cascade. The Gaussian distribution is illustrated by a solid line.  Note that except for the first scale ($\ell=0.0005$), they all have an exponential tails and do collapse with each other. 
b)  Extended Self-Similarity plots of the SFs on the range $0.0005<\ell<0.005$, corresponding to a wavenumber range $200<k<2000$. The solid line is a power-law fitting on this range by using a least square algorithm. For display clarity, these curves have been vertical shifted. }
    }\label{fig:ESS}
\end{figure*}

\begin{figure*}[!htb]
\centering
 \includegraphics[width=0.85\linewidth,clip]{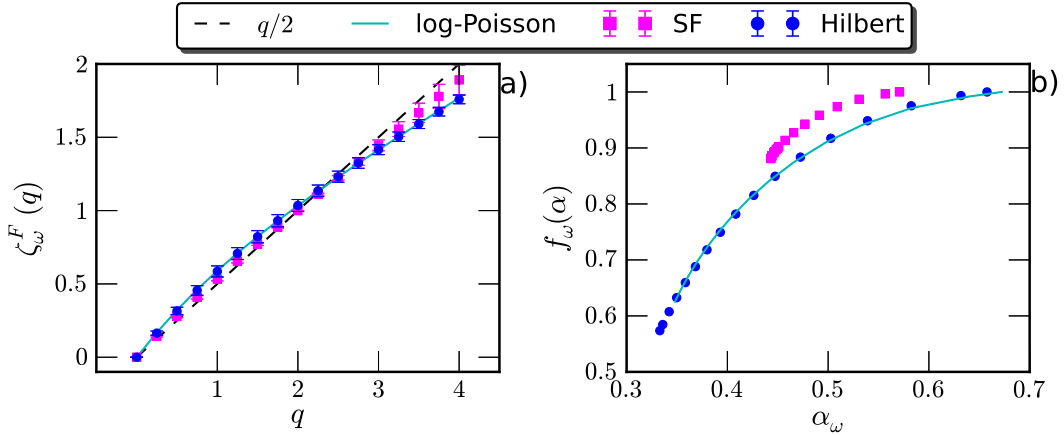}
  \caption{(Color online) \red{a) Comparison of the measured relative scaling exponents $\zeta_{\omega}^E(q)$ for the SFs ($\square$), the scaling exponent $\zeta_{\omega}^F(q)$ provided by the Hilbert method ($\ocircle$) and the  log-Poisson fitting (solid line). b) The corresponding singularity spectrum $f_{\omega}(\alpha)$. }
    }\label{fig:singularity}
\end{figure*}

\red{As we shown above that due to the large-scale structure influence the second-order SF fails to identify the  power-law behavior of the forward enstrophy cascade, see Fig. \ref{fig:psd}\,b).  Here we apply the Extended Self-Similarity (ESS) technique to extract the relative scaling exponents.\cite{Benzi1993PRE} Note that for the forward enstrophy cascade, the
second-order Hilbert  moments provides a scaling exponent $\zeta^F_{\omega}(2)\simeq1$  and the Fourier power spectrum provides $\beta\simeq2$. Therefore,  we define the ESS of the SFs by using the second-order SFs, i.e.,
\begin{equation}
S_q(\ell)\sim \left( S_q(\ell)\right)^{\zeta^E_{\omega}(q)}
\end{equation}
in which $S_q(\ell)=\langle  \vert\Delta_{\ell}\omega\vert^q \rangle$, and $\zeta^E_{\omega}(q)$ is the ESS scaling exponent.
Figure \ref{fig:ESS}\,a) shows the measured pdf of the vorticity increment for different separation scales in  the forward enstrophy cascade. Except for  $\ell=0.005$, all the pdfs have an exponential tail and  almost collapse with each other. 
Figure \ref{fig:ESS}\,b) shows the ESS plots on the range $0.0005<\ell<0.005$, 
corresponding to a wavenumber range $200<k<2000$, which is the scaling range 
of the forward enstrophy cascade predicted by the Hilbert method. Power-law 
behavior is observed for all $q$ we considered here. The ESS scaling is  estimated on this range by 
using a least square fitting algorithm.
  Figure \ref{fig:singularity}\,a) shows the 
measured ESS scaling exponent $\zeta_{\omega}^E(q)$ ($\square$), in which the 
errorbar indicates the 95\% fitting confidence. For comparison, the scaling exponent 
provided by the Hilbert method ($\ocircle$) is also shown. Graphically, the measured $\zeta^E_{\omega}(q)$ indicates a less intermittent vorticity field. To emphasize this point, the singularity spectrum $f_{\omega}(\alpha)$ is then calculated, i.e.,
  \begin{equation}
\alpha_{\omega}=\zeta_{\omega}(q)\rq{},\,\quad f_{\omega}(\alpha)=\min\left\{\alpha_{\omega} q -\zeta_{\omega}(q)+1 \right\}
\end{equation}
in which the scaling exponents $\zeta_{\omega}(q)$ could be either scaling exponent from Hilbert method or one from the SFs. Figure\,\ref{fig:singularity}\,b) shows the measured $f_{\omega}(\alpha)$ for the forward enstrophy cascade, in which 
the log-Poisson fitting is illustrated by a solid line.  It confirms again that the SFs predicts a less intermittent vorticity field.
}	

\subsection{Discussion}
\red{Several works have been reported  for the forward enstrophy cascade that the Fourier power spectrum of vorticity field possesses a \lq{}$-2$\rq{} power-law
behavior rather than \lq{}$-1$\rq{}
 one required by Kraichnan\rq{}s theory, see Eq.\,\eqref{eq:vorticity}. However, there is no theory explanation for this contradiction. Specifically for the Hilbert method, we  note that the second-order statistics $\mathcal{L}_2(k)$ provides $\zeta_{\omega}^F(2)\simeq 1.0$  (resp. \lq{}$ -2.0$\rq{} for the Hilbert energy spectrum) rather than $0$ required by the dimensional argument.  
The high-order scaling exponents $\zeta_{\omega}(q)$  provided by the Hilbert method for both the inverse energy cascade and forward enstrophy cascade disagree with the Kraichnan\rq{}s theory prediction, see  Eq.\,\eqref{eq:HQ}.\cite{Kraichnan1967PoF,Falkovich1994PRE} 
Furthermore, they do not agree with the logarithmic correction  theory  either.\cite{Falkovich1994PRE}
  It suggests that  a new theory  is required in future to interpret the vorticity field of 2D turbulence to take into account not only this inconsistence but also the intermittent effect.}

We emphasize here that the scaling property of the forward enstrophy cascade might  
depend on the Ekman friction, the parameter $\alpha$.
\cite{Boffetta2002PRE,Biferale2003PoF,Boffetta2007JFM} Therefore, more data sets should be certainly investigated in future to see whether the scaling behavior reported in this work is universal for different  $\alpha$.

\section{Conclusion}

In summary,  we have applied   the Hilbert methodology to  the 
vorticity  field
obtained from a high resolution DNS of 2D turbulence.  A dual-cascade   with almost one decade scales for the inverse energy cascade and forward enstrophy cascade is
identified. The scaling exponents $\zeta_{\omega}(q)$ are extracted. In the inverse energy cascade, the pdf  $p(C,k)$ is collapsed with each other with an exponential tail. This indicates a nonintermittent  cascade process, which is confirmed by 
the measured scaling exponent $\zeta_{\omega}^I(q)=-q/3$.  In the forward enstrophy cascade,  
the measured pdf $p(C,k)$ also have  an  exponential tail. However, due to the different shape of the core part, they can not collapse with each other, indicating an intermittent forward enstrophy cascade.  The measured scaling exponent $\zeta_{\omega}^F(q)$ is nonlinear with $q$ when $q\le 2$, showing 
intermittency.  A log-Poisson fitting, i.e., $\zeta_{\omega}^F(q)=q/3+0.45(1-0.43^q)$, is thus proposed to characterize the measured $\zeta_{\omega}^F(q)$.

\begin{acknowledgments}
 This work is sponsored by the National Natural Science Foundation of China under Grant (Nos. 11072139, 11032007, 11272196, 11202122 and 11332006) , \lq{}Pu Jiang\rq{} project of Shanghai (No. 12PJ1403500) and  the 
 Shanghai Program for Innovative Research Team in Universities. 
We thank Professor G. Boffetta for providing us the DNS data, which  is freely available from the
iCFDdatabase.\footnote{{http://cfd.cineca.it}}
 Y.H. thanks Prof. L. Biferale and Prof. G.  Falkovich for useful discussions and comments. 
We thank
 Dr Gabriel Rilling and Professor Patrick Flandrin from laboratoire de Physique, 
CNRS \& ENS
Lyon (France)  for sharing their EMD code with us. \footnote{{http://perso.ens-lyon.fr/patrick.flandrin/emd.html}} 
We also thank the anonymous referees for their useful comments and suggestions.
\end{acknowledgments}


\begin{thebibliography}{48}%
\makeatletter
\providecommand \@ifxundefined [1]{%
 \@ifx{#1\undefined}
}%
\providecommand \@ifnum [1]{%
 \ifnum #1\expandafter \@firstoftwo
 \else \expandafter \@secondoftwo
 \fi
}%
\providecommand \@ifx [1]{%
 \ifx #1\expandafter \@firstoftwo
 \else \expandafter \@secondoftwo
 \fi
}%
\providecommand \natexlab [1]{#1}%
\providecommand \enquote  [1]{``#1''}%
\providecommand \bibnamefont  [1]{#1}%
\providecommand \bibfnamefont [1]{#1}%
\providecommand \citenamefont [1]{#1}%
\providecommand \href@noop [0]{\@secondoftwo}%
\providecommand \href [0]{\begingroup \@sanitize@url \@href}%
\providecommand \@href[1]{\@@startlink{#1}\@@href}%
\providecommand \@@href[1]{\endgroup#1\@@endlink}%
\providecommand \@sanitize@url [0]{\catcode `\\12\catcode `\$12\catcode
  `\&12\catcode `\#12\catcode `\^12\catcode `\_12\catcode `\%12\relax}%
\providecommand \@@startlink[1]{}%
\providecommand \@@endlink[0]{}%
\providecommand \url  [0]{\begingroup\@sanitize@url \@url }%
\providecommand \@url [1]{\endgroup\@href {#1}{\urlprefix }}%
\providecommand \urlprefix  [0]{URL }%
\providecommand \Eprint [0]{\href }%
\providecommand \doibase [0]{http://dx.doi.org/}%
\providecommand \selectlanguage [0]{\@gobble}%
\providecommand \bibinfo  [0]{\@secondoftwo}%
\providecommand \bibfield  [0]{\@secondoftwo}%
\providecommand \translation [1]{[#1]}%
\providecommand \BibitemOpen [0]{}%
\providecommand \bibitemStop [0]{}%
\providecommand \bibitemNoStop [0]{.\EOS\space}%
\providecommand \EOS [0]{\spacefactor3000\relax}%
\providecommand \BibitemShut  [1]{\csname bibitem#1\endcsname}%
\let\auto@bib@innerbib\@empty
\bibitem [{\citenamefont {Kraichnan}\ and\ \citenamefont
  {Montgomery}(1980)}]{Kraichnan1980RepProgPhys}%
  \BibitemOpen
  \bibfield  {author} {\bibinfo {author} {\bibfnamefont {R.}~\bibnamefont
  {Kraichnan}}\ and\ \bibinfo {author} {\bibfnamefont {D.}~\bibnamefont
  {Montgomery}},\ }\bibfield  {title} {\enquote {\bibinfo {title}
  {Two-dimensional turbulence},}\ }\href@noop {} {\bibfield  {journal}
  {\bibinfo  {journal} {Rep.Prog. Phys.}\ }\textbf {\bibinfo {volume} {43}},\
  \bibinfo {pages} {547} (\bibinfo {year} {1980})}\BibitemShut {NoStop}%
\bibitem [{\citenamefont {Tabeling}(2002)}]{Tabeling2002PhysRep}%
  \BibitemOpen
  \bibfield  {author} {\bibinfo {author} {\bibfnamefont {P.}~\bibnamefont
  {Tabeling}},\ }\bibfield  {title} {\enquote {\bibinfo {title}
  {Two-dimensional turbulence: a physicist approach},}\ }\href@noop {}
  {\bibfield  {journal} {\bibinfo  {journal} {Phys. Rep.}\ }\textbf {\bibinfo
  {volume} {362}},\ \bibinfo {pages} {1--62} (\bibinfo {year}
  {2002})}\BibitemShut {NoStop}%
\bibitem [{\citenamefont {Kellay}\ and\ \citenamefont
  {Goldburg}(2002)}]{Kellay2002RepProgPhys}%
  \BibitemOpen
  \bibfield  {author} {\bibinfo {author} {\bibfnamefont {H.}~\bibnamefont
  {Kellay}}\ and\ \bibinfo {author} {\bibfnamefont {W.}~\bibnamefont
  {Goldburg}},\ }\bibfield  {title} {\enquote {\bibinfo {title}
  {Two-dimensional turbulence: a review of some recent experiments},}\
  }\href@noop {} {\bibfield  {journal} {\bibinfo  {journal} {Rep. Prog. Phys.}\
  }\textbf {\bibinfo {volume} {65}},\ \bibinfo {pages} {845} (\bibinfo {year}
  {2002})}\BibitemShut {NoStop}%
\bibitem [{\citenamefont {Boffetta}\ and\ \citenamefont
  {Ecke}(2012)}]{Boffetta2012ARFM}%
  \BibitemOpen
  \bibfield  {author} {\bibinfo {author} {\bibfnamefont {G.}~\bibnamefont
  {Boffetta}}\ and\ \bibinfo {author} {\bibfnamefont {R.}~\bibnamefont
  {Ecke}},\ }\bibfield  {title} {\enquote {\bibinfo {title} {Two-dimensional
  turbulence},}\ }\href@noop {} {\bibfield  {journal} {\bibinfo  {journal}
  {Annu. Rev. Fluid Mech.}\ }\textbf {\bibinfo {volume} {44}},\ \bibinfo {pages}
  {427--51} (\bibinfo {year} {2012})}\BibitemShut {NoStop}%
\bibitem [{\citenamefont {Bouchet}\ and\ \citenamefont
  {Venaille}(2012)}]{Bouchet2012PhysRep}%
  \BibitemOpen
  \bibfield  {author} {\bibinfo {author} {\bibfnamefont {F.}~\bibnamefont
  {Bouchet}}\ and\ \bibinfo {author} {\bibfnamefont {A.}~\bibnamefont
  {Venaille}},\ }\bibfield  {title} {\enquote {\bibinfo {title} {Statistical
  mechanics of two-dimensional and geophysical flows},}\ }\href@noop {}
  {\bibfield  {journal} {\bibinfo  {journal} {Phys. Rep.}\ }\textbf {\bibinfo
  {volume} {515}},\ \bibinfo {pages} {227--95} (\bibinfo {year}
  {2012})}\BibitemShut {NoStop}%
\bibitem [{\citenamefont {Irion}(1999)}]{Irion1999Science}%
  \BibitemOpen
  \bibfield  {author} {\bibinfo {author} {\bibfnamefont {R.}~\bibnamefont
  {Irion}},\ }\bibfield  {title} {\enquote {\bibinfo {title} {Soap films reveal
  whirling worlds of turbulence},}\ }\href@noop {} {\bibfield  {journal}
  {\bibinfo  {journal} {Science}\ }\textbf {\bibinfo {volume} {284}},\ \bibinfo
  {pages} {1609--1610} (\bibinfo {year} {1999})}\BibitemShut {NoStop}%
\bibitem [{\citenamefont {Falkovich}\ and\ \citenamefont
  {Lebedev}(1994)}]{Falkovich1994PRE}%
  \BibitemOpen
  \bibfield  {author} {\bibinfo {author} {\bibfnamefont {G.}~\bibnamefont
  {Falkovich}}\ and\ \bibinfo {author} {\bibfnamefont {V.}~\bibnamefont
  {Lebedev}},\ }\bibfield  {title} {\enquote {\bibinfo {title} {Universal
  direct cascade in two-dimensional turbulence},}\ }\href@noop {} {\bibfield
  {journal} {\bibinfo  {journal} {Phys. Rev. E}\ }\textbf {\bibinfo {volume}
  {50}},\ \bibinfo {pages} {3883} (\bibinfo {year} {1994})}\BibitemShut
  {NoStop}%
\bibitem [{\citenamefont {Chen}\ \emph {et~al.}(2003)\citenamefont {Chen},
  \citenamefont {Ecke}, \citenamefont {Eyink}, \citenamefont {Wang},\ and\
  \citenamefont {Xiao}}]{Chen2003PRL}%
  \BibitemOpen
  \bibfield  {author} {\bibinfo {author} {\bibfnamefont {S.}~\bibnamefont
  {Chen}}, \bibinfo {author} {\bibfnamefont {R.}~\bibnamefont {Ecke}}, \bibinfo
  {author} {\bibfnamefont {G.}~\bibnamefont {Eyink}}, \bibinfo {author}
  {\bibfnamefont {X.}~\bibnamefont {Wang}}, \ and\ \bibinfo {author}
  {\bibfnamefont {Z.}~\bibnamefont {Xiao}},\ }\bibfield  {title} {\enquote
  {\bibinfo {title} {Physical mechanism of the two-dimensional enstrophy
  cascade},}\ }\href@noop {} {\bibfield  {journal} {\bibinfo  {journal} {Phys.
  Rev. Lett.}\ }\textbf {\bibinfo {volume} {91}},\ \bibinfo {pages} {214501}
  (\bibinfo {year} {2003})}\BibitemShut {NoStop}%
\bibitem [{\citenamefont {Chen}\ \emph {et~al.}(2006)\citenamefont {Chen},
  \citenamefont {Ecke}, \citenamefont {Eyink}, \citenamefont {Rivera},
  \citenamefont {Wan},\ and\ \citenamefont {Xiao}}]{Chen2006PRL}%
  \BibitemOpen
  \bibfield  {author} {\bibinfo {author} {\bibfnamefont {S.}~\bibnamefont
  {Chen}}, \bibinfo {author} {\bibfnamefont {R.}~\bibnamefont {Ecke}}, \bibinfo
  {author} {\bibfnamefont {G.}~\bibnamefont {Eyink}}, \bibinfo {author}
  {\bibfnamefont {M.}~\bibnamefont {Rivera}}, \bibinfo {author} {\bibfnamefont
  {M.}~\bibnamefont {Wan}}, \ and\ \bibinfo {author} {\bibfnamefont
  {Z.}~\bibnamefont {Xiao}},\ }\bibfield  {title} {\enquote {\bibinfo {title}
  {Physical mechanism of the two-dimensional inverse energy cascade},}\
  }\href@noop {} {\bibfield  {journal} {\bibinfo  {journal} {Phys. Rev. Lett.}\
  }\textbf {\bibinfo {volume} {96}},\ \bibinfo {pages} {84502} (\bibinfo {year}
  {2006})}\BibitemShut {NoStop}%
\bibitem [{\citenamefont {Boffetta}\ and\ \citenamefont
  {Musacchio}(2010)}]{Boffetta2010PRE}%
  \BibitemOpen
  \bibfield  {author} {\bibinfo {author} {\bibfnamefont {G.}~\bibnamefont
  {Boffetta}}\ and\ \bibinfo {author} {\bibfnamefont {S.}~\bibnamefont
  {Musacchio}},\ }\bibfield  {title} {\enquote {\bibinfo {title} {Evidence for
  the double cascade scenario in two-dimensional turbulence},}\ }\href@noop {}
  {\bibfield  {journal} {\bibinfo  {journal} {Phys. Rev. E}\ }\textbf {\bibinfo
  {volume} {82}},\ \bibinfo {pages} {016307} (\bibinfo {year}
  {2010})}\BibitemShut {NoStop}%
\bibitem [{\citenamefont {Alexakis}\ and\ \citenamefont
  {Doering}(2006)}]{Alexakis2006PLA}%
  \BibitemOpen
  \bibfield  {author} {\bibinfo {author} {\bibfnamefont {A.}~\bibnamefont
  {Alexakis}}\ and\ \bibinfo {author} {\bibfnamefont {C.}~\bibnamefont
  {Doering}},\ }\bibfield  {title} {\enquote {\bibinfo {title} {Energy and
  enstrophy dissipation in steady state 2D turbulence},}\ }\href@noop {}
  {\bibfield  {journal} {\bibinfo  {journal} {Phys. Lett. A}\ }\textbf
  {\bibinfo {volume} {359}},\ \bibinfo {pages} {652--657} (\bibinfo {year}
  {2006})}\BibitemShut {NoStop}%
\bibitem [{\citenamefont {Xia}\ \emph {et~al.}(2011)\citenamefont {Xia},
  \citenamefont {Byrne}, \citenamefont {Falkovich},\ and\ \citenamefont
  {Shats}}]{Xia2011NatPhys}%
  \BibitemOpen
  \bibfield  {author} {\bibinfo {author} {\bibfnamefont {H.}~\bibnamefont
  {Xia}}, \bibinfo {author} {\bibfnamefont {D.}~\bibnamefont {Byrne}}, \bibinfo
  {author} {\bibfnamefont {G.}~\bibnamefont {Falkovich}}, \ and\ \bibinfo
  {author} {\bibfnamefont {M.}~\bibnamefont {Shats}},\ }\bibfield  {title}
  {\enquote {\bibinfo {title} {Upscale energy transfer in thick turbulent fluid
  layers},}\ }\href@noop {} {\bibfield  {journal} {\bibinfo  {journal} {Nature
  Phys.}\ }\textbf {\bibinfo {volume} {7}},\ \bibinfo {pages} {321--324}
  (\bibinfo {year} {2011})}\BibitemShut {NoStop}%
\bibitem [{\citenamefont {Xia}\ \emph {et~al.}(2008)\citenamefont {Xia},
  \citenamefont {Punzmann}, \citenamefont {Falkovich},\ and\ \citenamefont
  {Shats}}]{Xia2008PRL}%
  \BibitemOpen
  \bibfield  {author} {\bibinfo {author} {\bibfnamefont {H.}~\bibnamefont
  {Xia}}, \bibinfo {author} {\bibfnamefont {H.}~\bibnamefont {Punzmann}},
  \bibinfo {author} {\bibfnamefont {G.}~\bibnamefont {Falkovich}}, \ and\
  \bibinfo {author} {\bibfnamefont {M.}~\bibnamefont {Shats}},\ }\bibfield
  {title} {\enquote {\bibinfo {title} {Turbulence-condensate interaction in two
  dimensions},}\ }\href@noop {} {\bibfield  {journal} {\bibinfo  {journal}
  {Phys. Rev. Lett.}\ }\textbf {\bibinfo {volume} {101}},\ \bibinfo {pages}
  {194504} (\bibinfo {year} {2008})}\BibitemShut {NoStop}%
\bibitem [{\citenamefont {Tran}\ \emph {et~al.}(2010)\citenamefont {Tran},
  \citenamefont {Chakraborty}, \citenamefont {Guttenberg}, \citenamefont
  {Prescott}, \citenamefont {Kellay}, \citenamefont {Goldburg}, \citenamefont
  {Goldenfeld},\ and\ \citenamefont {Gioia}}]{Tran2010NaturePhys}%
  \BibitemOpen
  \bibfield  {author} {\bibinfo {author} {\bibfnamefont {T.}~\bibnamefont
  {Tran}}, \bibinfo {author} {\bibfnamefont {P.}~\bibnamefont {Chakraborty}},
  \bibinfo {author} {\bibfnamefont {N.}~\bibnamefont {Guttenberg}}, \bibinfo
  {author} {\bibfnamefont {A.}~\bibnamefont {Prescott}}, \bibinfo {author}
  {\bibfnamefont {H.}~\bibnamefont {Kellay}}, \bibinfo {author} {\bibfnamefont
  {W.}~\bibnamefont {Goldburg}}, \bibinfo {author} {\bibfnamefont
  {N.}~\bibnamefont {Goldenfeld}}, \ and\ \bibinfo {author} {\bibfnamefont
  {G.}~\bibnamefont {Gioia}},\ }\bibfield  {title} {\enquote {\bibinfo {title}
  {Macroscopic effects of the spectral structure in turbulent flows},}\
  }\href@noop {} {\bibfield  {journal} {\bibinfo  {journal} {Nature Phys.}\
  }\textbf {\bibinfo {volume} {6}},\ \bibinfo {pages} {438--441} (\bibinfo
  {year} {2010})}\BibitemShut {NoStop}%
\bibitem [{\citenamefont {Kelley}\ and\ \citenamefont
  {Ouellette}(2011)}]{Kelley2011PoF}%
  \BibitemOpen
  \bibfield  {author} {\bibinfo {author} {\bibfnamefont {D.}~\bibnamefont
  {Kelley}}\ and\ \bibinfo {author} {\bibfnamefont {N.}~\bibnamefont
  {Ouellette}},\ }\bibfield  {title} {\enquote {\bibinfo {title}
  {Spatiotemporal persistence of spectral fluxes in two-dimensional weak
  turbulence},}\ }\href@noop {} {\bibfield  {journal} {\bibinfo  {journal}
  {Phys. Fluids}\ }\textbf {\bibinfo {volume} {23}},\ \bibinfo {pages}
  {115101--115101} (\bibinfo {year} {2011})}\BibitemShut {NoStop}%
\bibitem [{\citenamefont {Merrifield}, \citenamefont {Kelley},\ and\
  \citenamefont {Ouellette}(2010)}]{Merrifield2010PRL}%
  \BibitemOpen
  \bibfield  {author} {\bibinfo {author} {\bibfnamefont {S.}~\bibnamefont
  {Merrifield}}, \bibinfo {author} {\bibfnamefont {D.}~\bibnamefont {Kelley}},
  \ and\ \bibinfo {author} {\bibfnamefont {N.}~\bibnamefont {Ouellette}},\
  }\bibfield  {title} {\enquote {\bibinfo {title} {Scale-dependent statistical
  geometry in two-dimensional flow},}\ }\href@noop {} {\bibfield  {journal}
  {\bibinfo  {journal} {Phys. Rev. Lett.}\ }\textbf {\bibinfo {volume} {104}},\
  \bibinfo {pages} {254501} (\bibinfo {year} {2010})}\BibitemShut {NoStop}%
\bibitem [{\citenamefont {Celani}, \citenamefont {Musacchio},\ and\
  \citenamefont {Vincenzi}(2010)}]{Celani2010PRL}%
  \BibitemOpen
  \bibfield  {author} {\bibinfo {author} {\bibfnamefont {A.}~\bibnamefont
  {Celani}}, \bibinfo {author} {\bibfnamefont {S.}~\bibnamefont {Musacchio}}, \
  and\ \bibinfo {author} {\bibfnamefont {D.}~\bibnamefont {Vincenzi}},\
  }\bibfield  {title} {\enquote {\bibinfo {title} {Turbulence in more than two
  and less than three dimensions},}\ }\href@noop {} {\bibfield  {journal}
  {\bibinfo  {journal} {Phys. Rev. Lett.}\ }\textbf {\bibinfo {volume} {104}},\
  \bibinfo {pages} {184506} (\bibinfo {year} {2010})}\BibitemShut {NoStop}%
\bibitem [{\citenamefont {Khurana}\ and\ \citenamefont
  {Ouellette}(2012)}]{Khurana2012PoF}%
  \BibitemOpen
  \bibfield  {author} {\bibinfo {author} {\bibfnamefont {N.}~\bibnamefont
  {Khurana}}\ and\ \bibinfo {author} {\bibfnamefont {N.}~\bibnamefont
  {Ouellette}},\ }\bibfield  {title} {\enquote {\bibinfo {title} {Interactions
  between active particles and dynamical structures in chaotic flow},}\
  }\href@noop {} {\bibfield  {journal} {\bibinfo  {journal} {Phys. Fluids}\
  }\textbf {\bibinfo {volume} {24}},\ \bibinfo {pages} {091902--091902}
  (\bibinfo {year} {2012})}\BibitemShut {NoStop}%
\bibitem [{\citenamefont {Van~Heijst}\ and\ \citenamefont
  {Clercx}(2009)}]{Heijst2009ARFM}%
  \BibitemOpen
  \bibfield  {author} {\bibinfo {author} {\bibfnamefont {G.}~\bibnamefont
  {Van~Heijst}}\ and\ \bibinfo {author} {\bibfnamefont {H.}~\bibnamefont
  {Clercx}},\ }\bibfield  {title} {\enquote {\bibinfo {title} {Laboratory
  modeling of geophysical vortices},}\ }\href@noop {} {\bibfield  {journal}
  {\bibinfo  {journal} {Annu. Rev. Fluid Mech.}\ }\textbf {\bibinfo {volume}
  {41}},\ \bibinfo {pages} {143--164} (\bibinfo {year} {2009})}\BibitemShut
  {NoStop}%
\bibitem [{\citenamefont {Boffetta}\ \emph {et~al.}(2002)\citenamefont
  {Boffetta}, \citenamefont {Celani}, \citenamefont {Musacchio},\ and\
  \citenamefont {Vergassola}}]{Boffetta2002PRE}%
  \BibitemOpen
  \bibfield  {author} {\bibinfo {author} {\bibfnamefont {G.}~\bibnamefont
  {Boffetta}}, \bibinfo {author} {\bibfnamefont {A.}~\bibnamefont {Celani}},
  \bibinfo {author} {\bibfnamefont {S.}~\bibnamefont {Musacchio}}, \ and\
  \bibinfo {author} {\bibfnamefont {M.}~\bibnamefont {Vergassola}},\ }\bibfield
   {title} {\enquote {\bibinfo {title} {Intermittency in two-dimensional
  Ekman-Navier-Stokes turbulence},}\ }\href@noop {} {\bibfield  {journal}
  {\bibinfo  {journal} {Phys. Rev. E}\ }\textbf {\bibinfo {volume} {66}},\
  \bibinfo {pages} {026304} (\bibinfo {year} {2002})}\BibitemShut {NoStop}%
\bibitem [{\citenamefont {Boffetta}(2007)}]{Boffetta2007JFM}%
  \BibitemOpen
  \bibfield  {author} {\bibinfo {author} {\bibfnamefont {G.}~\bibnamefont
  {Boffetta}},\ }\bibfield  {title} {\enquote {\bibinfo {title} {Energy and
  enstrophy fluxes in the double cascade of two-dimensional turbulence},}\
  }\href@noop {} {\bibfield  {journal} {\bibinfo  {journal} {J. Fluid Mech.}\
  }\textbf {\bibinfo {volume} {589}},\ \bibinfo {pages} {253--260} (\bibinfo
  {year} {2007})}\BibitemShut {NoStop}%
\bibitem [{\citenamefont {Kraichnan}(1967)}]{Kraichnan1967PoF}%
  \BibitemOpen
  \bibfield  {author} {\bibinfo {author} {\bibfnamefont {R.}~\bibnamefont
  {Kraichnan}},\ }\bibfield  {title} {\enquote {\bibinfo {title} {Inertial
  ranges in two-dimensional turbulence},}\ }\href@noop {} {\bibfield  {journal}
  {\bibinfo  {journal} {Phys. Fluids}\ }\textbf {\bibinfo {volume} {10}},\
  \bibinfo {pages} {1417--1423} (\bibinfo {year} {1967})}\BibitemShut {NoStop}%
\bibitem [{\citenamefont {Frisch}(1995)}]{Frisch1995}%
  \BibitemOpen
  \bibfield  {author} {\bibinfo {author} {\bibfnamefont {U.}~\bibnamefont
  {Frisch}},\ }\href@noop {} {\emph {\bibinfo {title} {{Turbulence: the legacy
  of AN Kolmogorov}}}}\ (\bibinfo  {publisher} {Cambridge University Press},\
  \bibinfo {year} {1995})\BibitemShut {NoStop}%
\bibitem [{\citenamefont {Huang}\ \emph {et~al.}(2010)\citenamefont {Huang},
  \citenamefont {Schmitt}, \citenamefont {Lu}, \citenamefont {Fougairolles},
  \citenamefont {Gagne},\ and\ \citenamefont {Liu}}]{Huang2010PRE}%
  \BibitemOpen
  \bibfield  {author} {\bibinfo {author} {\bibfnamefont {Y.}~\bibnamefont
  {Huang}}, \bibinfo {author} {\bibfnamefont {F.}~\bibnamefont {Schmitt}},
  \bibinfo {author} {\bibfnamefont {Z.}~\bibnamefont {Lu}}, \bibinfo {author}
  {\bibfnamefont {P.}~\bibnamefont {Fougairolles}}, \bibinfo {author}
  {\bibfnamefont {Y.}~\bibnamefont {Gagne}}, \ and\ \bibinfo {author}
  {\bibfnamefont {Y.}~\bibnamefont {Liu}},\ }\bibfield  {title} {\enquote
  {\bibinfo {title} {{Second-order structure function in fully developed
  turbulence}},}\ }\href@noop {} {\bibfield  {journal} {\bibinfo  {journal}
  {Phys. Rev. E}\ }\textbf {\bibinfo {volume} {82}},\ \bibinfo {pages} {026319}
  (\bibinfo {year} {2010})}\BibitemShut {NoStop}%
\bibitem [{\citenamefont {Biferale}\ \emph {et~al.}(2003)\citenamefont
  {Biferale}, \citenamefont {Cencini}, \citenamefont {Lanotte},\ and\
  \citenamefont {Vergni}}]{Biferale2003PoF}%
  \BibitemOpen
  \bibfield  {author} {\bibinfo {author} {\bibfnamefont {L.}~\bibnamefont
  {Biferale}}, \bibinfo {author} {\bibfnamefont {M.}~\bibnamefont {Cencini}},
  \bibinfo {author} {\bibfnamefont {A.}~\bibnamefont {Lanotte}}, \ and\
  \bibinfo {author} {\bibfnamefont {D.}~\bibnamefont {Vergni}},\ }\bibfield
  {title} {\enquote {\bibinfo {title} {Inverse velocity statistics in
  two-dimensional turbulence},}\ }\href@noop {} {\bibfield  {journal} {\bibinfo
   {journal} {Phys. Fluids}\ }\textbf {\bibinfo {volume} {15}},\ \bibinfo
  {pages} {1012--1020} (\bibinfo {year} {2003})}\BibitemShut {NoStop}%
\bibitem [{\citenamefont {Kellay}, \citenamefont {Wu},\ and\ \citenamefont
  {Goldburg}(1998)}]{Kellay1998PRL}%
  \BibitemOpen
  \bibfield  {author} {\bibinfo {author} {\bibfnamefont {H.}~\bibnamefont
  {Kellay}}, \bibinfo {author} {\bibfnamefont {X.}~\bibnamefont {Wu}}, \ and\
  \bibinfo {author} {\bibfnamefont {W.}~\bibnamefont {Goldburg}},\ }\bibfield
  {title} {\enquote {\bibinfo {title} {Vorticity measurements in turbulent soap
  films},}\ }\href@noop {} {\bibfield  {journal} {\bibinfo  {journal} {Phys.
  Rev. Lett.}\ }\textbf {\bibinfo {volume} {80}},\ \bibinfo {pages} {277--280}
  (\bibinfo {year} {1998})}\BibitemShut {NoStop}%
\bibitem [{\citenamefont {Paret}, \citenamefont {Jullien},\ and\ \citenamefont
  {Tabeling}(1999)}]{Paret1999PRL}%
  \BibitemOpen
  \bibfield  {author} {\bibinfo {author} {\bibfnamefont {J.}~\bibnamefont
  {Paret}}, \bibinfo {author} {\bibfnamefont {M.}~\bibnamefont {Jullien}}, \
  and\ \bibinfo {author} {\bibfnamefont {P.}~\bibnamefont {Tabeling}},\
  }\bibfield  {title} {\enquote {\bibinfo {title} {Vorticity statistics in the
  two-dimensional enstrophy cascade},}\ }\href@noop {} {\bibfield  {journal}
  {\bibinfo  {journal} {Phys. Rev. Lett.}\ }\textbf {\bibinfo {volume} {83}},\
  \bibinfo {pages} {3418--3421} (\bibinfo {year} {1999})}\BibitemShut {NoStop}%
\bibitem [{\citenamefont {Nam}\ \emph {et~al.}(2000)\citenamefont {Nam},
  \citenamefont {Ott}, \citenamefont {Antonsen~Jr},\ and\ \citenamefont
  {Guzdar}}]{Nam2000PRL}%
  \BibitemOpen
  \bibfield  {author} {\bibinfo {author} {\bibfnamefont {K.}~\bibnamefont
  {Nam}}, \bibinfo {author} {\bibfnamefont {E.}~\bibnamefont {Ott}}, \bibinfo
  {author} {\bibfnamefont {T.}~\bibnamefont {Antonsen~Jr}}, \ and\ \bibinfo
  {author} {\bibfnamefont {P.}~\bibnamefont {Guzdar}},\ }\bibfield  {title}
  {\enquote {\bibinfo {title} {Lagrangian chaos and the effect of drag on the
  enstrophy cascade in two-dimensional turbulence},}\ }\href@noop {} {\bibfield
   {journal} {\bibinfo  {journal} {Phys. Rev. Lett.}\ }\textbf {\bibinfo
  {volume} {84}},\ \bibinfo {pages} {5134--5137} (\bibinfo {year}
  {2000})}\BibitemShut {NoStop}%
\bibitem [{\citenamefont {Bernard}(2000)}]{Bernard2000EPL}%
  \BibitemOpen
  \bibfield  {author} {\bibinfo {author} {\bibfnamefont {D.}~\bibnamefont
  {Bernard}},\ }\bibfield  {title} {\enquote {\bibinfo {title} {Influence of
  friction on the direct cascade of the 2D forced turbulence},}\ }\href@noop {}
  {\bibfield  {journal} {\bibinfo  {journal} {Europhys. Lett.}\
  }\textbf {\bibinfo {volume} {50}},\ \bibinfo {pages} {333} (\bibinfo {year}
  {2000})}\BibitemShut {NoStop}%
\bibitem [{\citenamefont {Tsang}\ \emph {et~al.}(2005)\citenamefont {Tsang},
  \citenamefont {Ott}, \citenamefont {Antonsen~Jr},\ and\ \citenamefont
  {Guzdar}}]{Tsang2005PRE}%
  \BibitemOpen
  \bibfield  {author} {\bibinfo {author} {\bibfnamefont {Y.}~\bibnamefont
  {Tsang}}, \bibinfo {author} {\bibfnamefont {E.}~\bibnamefont {Ott}}, \bibinfo
  {author} {\bibfnamefont {T.}~\bibnamefont {Antonsen~Jr}}, \ and\ \bibinfo
  {author} {\bibfnamefont {P.}~\bibnamefont {Guzdar}},\ }\bibfield  {title}
  {\enquote {\bibinfo {title} {Intermittency in two-dimensional turbulence with
  drag},}\ }\href@noop {} {\bibfield  {journal} {\bibinfo  {journal} {Phys.
  Rev. E}\ }\textbf {\bibinfo {volume} {71}},\ \bibinfo {pages} {066313}
  (\bibinfo {year} {2005})}\BibitemShut {NoStop}%
\bibitem [{\citenamefont {Falkovich}\ and\ \citenamefont
  {Lebedev}(2011)}]{Falkovich2011PRE}%
  \BibitemOpen
  \bibfield  {author} {\bibinfo {author} {\bibfnamefont {G.}~\bibnamefont
  {Falkovich}}\ and\ \bibinfo {author} {\bibfnamefont {V.}~\bibnamefont
  {Lebedev}},\ }\bibfield  {title} {\enquote {\bibinfo {title} {Vorticity
  statistics in the direct cascade of two-dimensional turbulence},}\
  }\href@noop {} {\bibfield  {journal} {\bibinfo  {journal} {Phys. Rev. E}\
  }\textbf {\bibinfo {volume} {83}},\ \bibinfo {pages} {045301} (\bibinfo
  {year} {2011})}\BibitemShut {NoStop}%
\bibitem [{\citenamefont {Falkovich}\ and\ \citenamefont
  {Sreenivasan}(2006)}]{Falkovich2006PhysToday}%
  \BibitemOpen
  \bibfield  {author} {\bibinfo {author} {\bibfnamefont {G.}~\bibnamefont
  {Falkovich}}\ and\ \bibinfo {author} {\bibfnamefont {K.~R.}\ \bibnamefont
  {Sreenivasan}},\ }\bibfield  {title} {\enquote {\bibinfo {title} {Lessons
  from hydrodynamic turbulence},}\ }\href@noop {} {\bibfield  {journal}
  {\bibinfo  {journal} {Phys. Today}\ }\textbf {\bibinfo {volume} {59}},\
  \bibinfo {pages} {43} (\bibinfo {year} {2006})}\BibitemShut {NoStop}%
\bibitem [{\citenamefont {Huang}\ \emph {et~al.}(2008)\citenamefont {Huang},
  \citenamefont {Schmitt}, \citenamefont {Lu},\ and\ \citenamefont
  {Liu}}]{Huang2008EPL}%
  \BibitemOpen
  \bibfield  {author} {\bibinfo {author} {\bibfnamefont {Y.}~\bibnamefont
  {Huang}}, \bibinfo {author} {\bibfnamefont {F.}~\bibnamefont {Schmitt}},
  \bibinfo {author} {\bibfnamefont {Z.}~\bibnamefont {Lu}}, \ and\ \bibinfo
  {author} {\bibfnamefont {Y.}~\bibnamefont {Liu}},\ }\bibfield  {title}
  {\enquote {\bibinfo {title} {An amplitude-frequency study of turbulent
  scaling intermittency using Hilbert spectral analysis},}\ }\href@noop {}
  {\bibfield  {journal} {\bibinfo  {journal} {Europhys. Lett.}\ }\textbf
  {\bibinfo {volume} {84}},\ \bibinfo {pages} {40010} (\bibinfo {year}
  {2008})}\BibitemShut {NoStop}%
\bibitem [{\citenamefont {Huang}\ \emph {et~al.}(2011)\citenamefont {Huang},
  \citenamefont {Schmitt}, \citenamefont {Hermand}, \citenamefont {Gagne},
  \citenamefont {Lu},\ and\ \citenamefont {Liu}}]{Huang2011PRE}%
  \BibitemOpen
  \bibfield  {author} {\bibinfo {author} {\bibfnamefont {Y.}~\bibnamefont
  {Huang}}, \bibinfo {author} {\bibfnamefont {F.}\ \bibnamefont {Schmitt}},
  \bibinfo {author} {\bibfnamefont {J.-P.}\ \bibnamefont {Hermand}}, \bibinfo
  {author} {\bibfnamefont {Y.}~\bibnamefont {Gagne}}, \bibinfo {author}
  {\bibfnamefont {Z.}~\bibnamefont {Lu}}, \ and\ \bibinfo {author}
  {\bibfnamefont {Y.}~\bibnamefont {Liu}},\ }\bibfield  {title} {\enquote
  {\bibinfo {title} {Arbitrary-order Hilbert spectral analysis for time series
  possessing scaling statistics: comparison study with detrended fluctuation
  analysis and wavelet leaders},}\ }\href@noop {} {\bibfield  {journal}
  {\bibinfo  {journal} {Phys. Rev. E}\ }\textbf {\bibinfo {volume} {84}},\
  \bibinfo {pages} {016208} (\bibinfo {year} {2011})}\BibitemShut {NoStop}%
\bibitem [{\citenamefont {Huang}\ \emph {et~al.}(1998)\citenamefont {Huang},
  \citenamefont {Shen}, \citenamefont {Long}, \citenamefont {Wu}, \citenamefont
  {Shih}, \citenamefont {Zheng}, \citenamefont {Yen}, \citenamefont {Tung},\
  and\ \citenamefont {Liu}}]{Huang1998EMD}%
  \BibitemOpen
  \bibfield  {author} {\bibinfo {author} {\bibfnamefont {N.~E.}\ \bibnamefont
  {Huang}}, \bibinfo {author} {\bibfnamefont {Z.}~\bibnamefont {Shen}},
  \bibinfo {author} {\bibfnamefont {S.~R.}\ \bibnamefont {Long}}, \bibinfo
  {author} {\bibfnamefont {M.~C.}\ \bibnamefont {Wu}}, \bibinfo {author}
  {\bibfnamefont {H.~H.}\ \bibnamefont {Shih}}, \bibinfo {author}
  {\bibfnamefont {Q.}~\bibnamefont {Zheng}}, \bibinfo {author} {\bibfnamefont
  {N.}~\bibnamefont {Yen}}, \bibinfo {author} {\bibfnamefont {C.~C.}\
  \bibnamefont {Tung}}, \ and\ \bibinfo {author} {\bibfnamefont {H.~H.}\
  \bibnamefont {Liu}},\ }\bibfield  {title} {\enquote {\bibinfo {title} {The
  empirical mode decomposition and the Hilbert spectrum for nonlinear and
  non-stationary time series analysis},}\ }\href@noop {} {\bibfield  {journal}
  {\bibinfo  {journal} {Proc. R. Soc. London, Ser. A}\ }\textbf {\bibinfo
  {volume} {454}},\ \bibinfo {pages} {903--995} (\bibinfo {year}
  {1998})}\BibitemShut {NoStop}%
\bibitem [{\citenamefont {Huang}, \citenamefont {Shen},\ and\ \citenamefont
  {Long}(1999)}]{Huang1999EMD}%
  \BibitemOpen
  \bibfield  {author} {\bibinfo {author} {\bibfnamefont {N.~E.}\ \bibnamefont
  {Huang}}, \bibinfo {author} {\bibfnamefont {Z.}~\bibnamefont {Shen}}, \ and\
  \bibinfo {author} {\bibfnamefont {S.~R.}\ \bibnamefont {Long}},\ }\bibfield
  {title} {\enquote {\bibinfo {title} {{A new view of nonlinear water waves:
  The Hilbert Spectrum }},}\ }\href@noop {} {\bibfield  {journal} {\bibinfo
  {journal} {Annu. Rev. Fluid Mech.}\ }\textbf {\bibinfo {volume} {31}},\
  \bibinfo {pages} {417--457} (\bibinfo {year} {1999})}\BibitemShut {NoStop}%
\bibitem [{\citenamefont {Flandrin}\ and\ \citenamefont
  {Gon{\c{c}}alv\`es}(2004)}]{Flandrin2004EMDa}%
  \BibitemOpen
  \bibfield  {author} {\bibinfo {author} {\bibfnamefont {P.}~\bibnamefont
  {Flandrin}}\ and\ \bibinfo {author} {\bibfnamefont {P.}~\bibnamefont
  {Gon{\c{c}}alv\`es}},\ }\bibfield  {title} {\enquote {\bibinfo {title}
  {{Empirical mode decompositions as data-driven Wavelet-like expansions}},}\
  }\href@noop {} {\bibfield  {journal} {\bibinfo  {journal} {Int. J. Wavelets,
  Multires. Info. Proc.}\ }\textbf {\bibinfo {volume} {2}},\ \bibinfo {pages}
  {477--496} (\bibinfo {year} {2004})}\BibitemShut {NoStop}%
\bibitem [{\citenamefont {Rilling}, \citenamefont {Flandrin},\ and\
  \citenamefont {Gon\c{c}alv\`es}(2003)}]{Rilling2003EMD}%
  \BibitemOpen
  \bibfield  {author} {\bibinfo {author} {\bibfnamefont {G.}~\bibnamefont
  {Rilling}}, \bibinfo {author} {\bibfnamefont {P.}~\bibnamefont {Flandrin}}, \
  and\ \bibinfo {author} {\bibfnamefont {P.}~\bibnamefont {Gon\c{c}alv\`es}},\
  }\bibfield  {title} {\enquote {\bibinfo {title} {{On empirical mode
  decomposition and its algorithms}},}\ }\href@noop {} {\bibfield  {journal}
  {\bibinfo  {journal} {IEEE-EURASIP Workshop on Nonlinear Signal and Image
  Processing}\ } (\bibinfo {year} {2003})}\BibitemShut {NoStop}%
\bibitem [{\citenamefont {Huang}(2009)}]{Huang2009PHD}%
  \BibitemOpen
  \bibfield  {author} {\bibinfo {author} {\bibfnamefont {Y.}~\bibnamefont
  {Huang}},\ }\emph {\bibinfo {title} {Arbitrary-Order Hilbert Spectral
  Analysis: Definition and Application to fully developed turbulence and
  environmental time series}},\ \href@noop {} {Ph.D. thesis},\ \bibinfo
  {school} {Universit\'e des Sciences et Technologies de Lille - Lille 1,
  France \& Shanghai University, China} (\bibinfo {year} {2009})\BibitemShut
  {NoStop}%
\bibitem [{\citenamefont {Cohen}(1995)}]{Cohen1995}%
  \BibitemOpen
  \bibfield  {author} {\bibinfo {author} {\bibfnamefont {L.}~\bibnamefont
  {Cohen}},\ }\href@noop {} {\emph {\bibinfo {title} {{Time-frequency
  analysis}}}}\ (\bibinfo  {publisher} {Prentice Hall PTR Englewood Cliffs,
  NJ},\ \bibinfo {year} {1995})\BibitemShut {NoStop}%
\bibitem [{\citenamefont {Flandrin}(1998)}]{Flandrin1998}%
  \BibitemOpen
  \bibfield  {author} {\bibinfo {author} {\bibfnamefont {P.}~\bibnamefont
  {Flandrin}},\ }\href@noop {} {\emph {\bibinfo {title}
  {{Time-frequency/time-scale analysis}}}}\ (\bibinfo  {publisher} {Academic
  Press},\ \bibinfo {year} {1998})\BibitemShut {NoStop}%
\bibitem [{\citenamefont {Huang}\ \emph {et~al.}(2013)\citenamefont {Huang},
  \citenamefont {Biferale}, \citenamefont {Calzavarini}, \citenamefont {Sun},\
  and\ \citenamefont {Toschi}}]{Huang2013PRE}%
  \BibitemOpen
  \bibfield  {author} {\bibinfo {author} {\bibfnamefont {Y.}~\bibnamefont
  {Huang}}, \bibinfo {author} {\bibfnamefont {L.}~\bibnamefont {Biferale}},
  \bibinfo {author} {\bibfnamefont {E.}~\bibnamefont {Calzavarini}}, \bibinfo
  {author} {\bibfnamefont {C.}~\bibnamefont {Sun}}, \ and\ \bibinfo {author}
  {\bibfnamefont {F.}~\bibnamefont {Toschi}},\ }\bibfield  {title} {\enquote
  {\bibinfo {title} {Lagrangian single particle turbulent statistics through
  the Hilbert-Huang transforms},}\ }\href@noop {} {\bibfield  {journal}
  {\bibinfo  {journal} {Phys. Rev. E}\ }\textbf {\bibinfo {volume} {87}},\
  \bibinfo {pages} {041003(R)} (\bibinfo {year} {2013})}\BibitemShut {NoStop}%
\bibitem [{\citenamefont {Pasquero}\ and\ \citenamefont
  {Falkovich}(2002)}]{Pasquero2002PRE}%
  \BibitemOpen
  \bibfield  {author} {\bibinfo {author} {\bibfnamefont {C.}~\bibnamefont
  {Pasquero}}\ and\ \bibinfo {author} {\bibfnamefont {G.}~\bibnamefont
  {Falkovich}},\ }\bibfield  {title} {\enquote {\bibinfo {title} {Stationary
  spectrum of vorticity cascade in two-dimensional turbulence},}\ }\href@noop
  {} {\bibfield  {journal} {\bibinfo  {journal} {Phys. Rev. E}\ }\textbf
  {\bibinfo {volume} {65}},\ \bibinfo {pages} {56305} (\bibinfo {year}
  {2002})}\BibitemShut {NoStop}%
\bibitem [{\citenamefont {Davidson}\ and\ \citenamefont
  {Pearson}(2005)}]{Davidson2005PRL}%
  \BibitemOpen
  \bibfield  {author} {\bibinfo {author} {\bibfnamefont {P.~A.}\ \bibnamefont
  {Davidson}}\ and\ \bibinfo {author} {\bibfnamefont {B.~R.}\ \bibnamefont
  {Pearson}},\ }\bibfield  {title} {\enquote {\bibinfo {title} {Identifying
  turbulent energy distribution in real, rather than fourier, space},}\
  }\href@noop {} {\bibfield  {journal} {\bibinfo  {journal} {Phys. Rev. Lett.}\
  }\textbf {\bibinfo {volume} {95}},\ \bibinfo {pages} {214501} (\bibinfo
  {year} {2005})}\BibitemShut {NoStop}%
\bibitem [{\citenamefont {Toschi}\ \emph {et~al.}(2005)\citenamefont {Toschi},
  \citenamefont {Biferale}, \citenamefont {Boffetta}, \citenamefont {Celani},
  \citenamefont {Devenish},\ and\ \citenamefont {Lanotte}}]{Toschi2005JOT}%
  \BibitemOpen
  \bibfield  {author} {\bibinfo {author} {\bibfnamefont {F.}~\bibnamefont
  {Toschi}}, \bibinfo {author} {\bibfnamefont {L.}~\bibnamefont {Biferale}},
  \bibinfo {author} {\bibfnamefont {G.}~\bibnamefont {Boffetta}}, \bibinfo
  {author} {\bibfnamefont {A.}~\bibnamefont {Celani}}, \bibinfo {author}
  {\bibfnamefont {B.}~\bibnamefont {Devenish}}, \ and\ \bibinfo {author}
  {\bibfnamefont {A.}~\bibnamefont {Lanotte}},\ }\bibfield  {title} {\enquote
  {\bibinfo {title} {Acceleration and vortex filaments in turbulence},}\
  }\href@noop {} {\bibfield  {journal} {\bibinfo  {journal} {J. Turbu.}\
  }\textbf {\bibinfo {volume} {6}},\ \bibinfo {pages} {15} (\bibinfo {year}
  {2005})}\BibitemShut {NoStop}%
\bibitem [{\citenamefont {Benzi}\ \emph {et~al.}(1993)\citenamefont {Benzi},
  \citenamefont {Ciliberto}, \citenamefont {Tripiccione}, \citenamefont
  {Baudet}, \citenamefont {Massaioli},\ and\ \citenamefont
  {Succi}}]{Benzi1993PRE}%
  \BibitemOpen
  \bibfield  {author} {\bibinfo {author} {\bibfnamefont {R.}~\bibnamefont
  {Benzi}}, \bibinfo {author} {\bibfnamefont {S.}~\bibnamefont {Ciliberto}},
  \bibinfo {author} {\bibfnamefont {R.}~\bibnamefont {Tripiccione}}, \bibinfo
  {author} {\bibfnamefont {C.}~\bibnamefont {Baudet}}, \bibinfo {author}
  {\bibfnamefont {F.}~\bibnamefont {Massaioli}}, \ and\ \bibinfo {author}
  {\bibfnamefont {S.}~\bibnamefont {Succi}},\ }\bibfield  {title} {\enquote
  {\bibinfo {title} {{Extended self-similarity in turbulent flows}},}\
  }\href@noop {} {\bibfield  {journal} {\bibinfo  {journal} {Phys. Rev. E}\
  }\textbf {\bibinfo {volume} {48}},\ \bibinfo {pages} {29--32} (\bibinfo
  {year} {1993})}\BibitemShut {NoStop}%
\bibitem [{Note1()}]{Note1}%
  \BibitemOpen
  \bibinfo {note} {{http://cfd.cineca.it}}\BibitemShut {NoStop}%
\bibitem [{Note2()}]{Note2}%
  \BibitemOpen
  \bibinfo {note}
  {{http://perso.ens-lyon.fr/patrick.flandrin/emd.html}}\BibitemShut {NoStop}%
\end{thebibliography}
%

\end{document}